\documentclass[11pt,a4paper]{article}
\usepackage{jheppub}

\usepackage{bbm,amsmath,graphicx,amssymb}
\usepackage{epstopdf}
\usepackage{bm}
\usepackage{verbatim}
\usepackage{slashed}
\usepackage{multirow}

\newcommand{\bea}{\begin{eqnarray}}
\newcommand{\eea}{\end{eqnarray}}
\newcommand{\bean}{\begin{eqnarray*}}
\newcommand{\eean}{\end{eqnarray*}}

\def\Label#1{\label{#1}%
  \smash{\hbox to0pt{\raise1ex\hbox{\tiny[#1]}\hss}}}

\newcommand{\bC}{\mathbb C}
\newcommand{\Var}{\mathcal V}
\newcommand{\sH}{\mathcal H}
\newcommand{\sW}{\mathcal W}

\newtheorem{remark}{Remark}

\def\stamp{--- {\bf \today} --- {\bf \jobname.tex}}

\def\fs_#1{\mathfrak{s}(#1)}

\def\BE{\begin{equation}}
\def\EE{\end{equation}}
\def\spa#1.#2{\left\langle#1\,#2\right\rangle}
\def\spb#1.#2{\left[#1\,#2\right]}
\def\lor#1.#2{\left(#1\,#2\right)}

\newcommand\fverb{\setbox\fverbbox=\hbox\bgroup\verb}
\newcommand\fverbdo{\egroup\medskip\noindent%
            \fbox{\unhbox\fverbbox}\ }
\newcommand\fverbit{\egroup\item[\fbox{\unhbox\fverbbox}]}
\newbox\fverbbox

\title{Global Structure of Curves from Generalized Unitarity Cut of Three-loop Diagrams}

\author[a]{Jonathan D. Hauenstein,}
\author[b]{Rijun Huang,}
\author[a]{Dhagash Mehta,}
\author[c]{Yang Zhang}

\affiliation[a]{Department of Applied and Computational Mathematics and Statistics, University of Notre Dame, Notre Dame, IN 46556, USA}
\affiliation[b]{Institut de Physique Th\'eorique, CEA-Saclay, F--91191
Gif-sur-Yvette cedex, France}
\affiliation[c]{Niels Bohr International Academy and Discovery Center,
The Niels Bohr Institute, \\University of Copenhagen, Blegdamsvej 17, DK-2100 Copenhagen \O,
Denmark}
\emailAdd{hauenstein@nd.edu,~huang@nbi.dk,~dmehta@nd.edu,~zhang@nbi.dk}
\abstract{This paper studies the global structure of algebraic curves defined by generalized unitarity cut of four-dimensional three-loop
diagrams with eleven propagators.  The global structure is
a topological invariant that is characterized by the geometric genus
of the algebraic curve.  We use the Riemann-Hurwitz formula to compute
the geometric genus of algebraic curves with the help of
techniques involving convex hull polytopes and numerical algebraic geometry. Some interesting properties of genus for arbitrary loop orders are also explored where computing the genus serves as an initial step
for integral or integrand reduction of three-loop amplitudes
via an algebraic geometric approach.}


\begin{document}
%%%%%%%%%%%%%%%%%%%%%%%%%%
\maketitle

%%%%%%%%%%%%%%%%%%%%%
\section{Introduction}
%%%%%%%%%%%%%%%%%%%%%%%

Algebraic geometry has been introduced to the multi-loop amplitude
computation in recent years, responding to the demand of
Next-to-Leading-Order or Next-Next-to-Leading-order precision
correction for collider experiments. Many attempts have been taken
towards the purpose of implementing a systematic and automatic
algorithm for two-loop and three-loop amplitude computations, in the
language of complex algebraic geometry.

The basic idea of this approach is to generalize traditional
concepts, such as integral reduction
\cite{Brown:1952eu,Passarino:1978jh,'tHooft:1978xw,Stuart:1987tt,Stuart:1989de}
and unitarity cut
\cite{Landau:1959fi,Mandelstam:1958xc,Mandelstam:1959bc,Cutkosky:1960sp},
from one-loop to multi-loop amplitudes. It is well-known that
unitarity can be applied to the computation of one-loop amplitudes
from tree amplitudes \cite{Bern:1994zx,Bern:1994cg,Bern:1995db},
while it is possible now to compute the tree amplitudes very
efficiently via the Britto-Cachazo-Feng-Witten recursion relation
\cite{Britto:2004ap,Britto:2005fq}. Moreover, one-loop integrals can
be algebraically reduced to a linear combination of scalar integrals
in the integral basis. The integral basis is a set of a finite number of
integrals. For one-loop amplitudes, it only contains scalar box,
triangle, bubble integrals in four-dimension, additional scalar
pentagon integral in $D$-dimension, and tadpole integral for massive
internal momenta. A scalar integral is an integral whose numerator
of integrand is one. All one-loop integrals with a tensor structure
in the numerator can be reduced by, e.g., Passarino-Veltman
reduction \cite{Passarino:1978jh,'tHooft:1978xw}, to scalar integral
basis. Starting from a general integral for a Feynman diagram, we can
perform the reduction procedure and keep track of all kinematic
factors in each step. Then the coefficients of integral basis can be
obtained as a consequence of reduction procedure. However, there is
a simpler way of computing the coefficients from tree amplitudes, by
matching the discontinuity of integrals under unitarity cuts
\cite{Britto:2005ha,Anastasiou:2006jv,Anastasiou:2006gt} or
generalized unitarity cuts \cite{Britto:2004nc,Bern:1997sc}.
Assuming that the integral basis is already known, which is indeed
the case for one-loop amplitudes, we can formally expand an one-loop
integral as a linear combination of integrals in the basis with
unknown coefficients. The box, triangle, bubble integrals under
unitarity cuts apparently have different signatures, which can be
used to identify the kinematic factors of signatures as coefficients
of integral basis correspondingly. In the integrand level, the
coefficients can also be extracted by reduction methods
algebraically \cite{Ossola:2006us}, with the help of quadruple,
triple and double unitarity cuts. This has been extensively suited
with numerical implementations
\cite{Forde:2007mi,Ellis:2007br,Kilgore:2007qr,Giele:2008ve,Ossola:2008xq,Badger:2008cm}.

The difficulty of generalizing the one-loop algorithm to multi-loop
amplitudes is obvious. First of all, the integral basis is generally
unknown. It is known that the integral basis contains not only
scalar integrals, but also tensor structure of loop momentum in the
numerator. Even for some simple diagrams such as a four-point two-loop
double-box diagram where the integral basis is already known
\cite{Gluza:2010ws}, unitarity cut method is not directly applicable
for determining the coefficients of the integral basis.  Algebraic
geometric techniques are introduced to overcome these difficulties,
and they provide a new interpretation of unitarity cut for
multi-loop amplitudes. It is applied both to integrand reduction and
integral reduction.

For multi-loop amplitude computations in the language of algebraic
geometry, the concept of maximal unitarity cut is replaced by the
simultaneous solution of on-shell equations of propagators, instead
of delta function constraints. The on-shell equations form the basis
of an ideal, and the simultaneous solution set defines the variety of the
ideal. In the integrand level \cite{Zhang:2012ce,Mastrolia:2012an},
the Gr\"{o}bner basis of the ideal is used as divisors, and
polynomial division over these divisors provides a finite set of
algebraically independent monomials, which defines the integrand
basis of a given diagram, in principle, to any loop orders. The
primary decomposition method is applied to study the irreducible
components of the ideal and the variety, which is useful for
determining coefficients of integrand basis through branch-by-branch
polynomial fitting method. These algebraic geometry techniques have
already been applied to the study of integrand basis and structure
of varieties of all four-dimensional two-loop diagrams, and for
explicitly computing some two-loop amplitudes and three-loop
amplitudes
\cite{Badger:2012dp,Feng:2012bm,Kleiss:2012yv,Badger:2012dv,Mastrolia:2012wf,Mastrolia:2012du,Mastrolia:2013kca,Badger:2013gxa,vanDeurzen:2013saa}.

In the integral level, the Integration-by-Parts(IBP) method
\cite{Tkachov:1981wb,Chetyrkin:1981qh,Laporta:2000dc,Laporta:2001dd}
is a traditional way of determining the integral basis from the
integrand basis. Recently, an attempt of determining integral basis
by unitarity cut method and spinor integration technique has also been
presented \cite{Feng:2014nwa}. Determining the integral basis of
multi-loop amplitudes is a non-trivial problem, and one of the
bottlenecks is that the computation is time-consuming even with a
computer. Thus it deserves more studies at both theoretical
\cite{Henn:2013pwa,Caron-Huot:2014lda} and computational levels.
Once the integral basis is determined for a given diagram,
algebraic geometry can be applied to the computation of their
expansion coefficients \cite{Kosower:2011ty}. Again, this is
realized by considering a simple fact that integration of a delta
function $\int dz~\delta(z)$ in $\mathbb{R}$ is equivalent to a
contour integration $\oint {dz\over z}$ in $\mathbb{C}$ by Cauchy's
integral theorem. The latter is in fact the computation of residues
at poles surrounded by chosen contours. In order for it to be
applied to multi-loop amplitude computations, it should be
generalized to multivariate analytic functions, which leads to the
computation of multivariate residues at global poles. The global
poles are determined by the simultaneous solution of on-shell
equations of propagators, and it requires the study of ideal and
variety of on-shell equations. The coefficients of integral basis
are computed as a linear combination of contour integrations at some
chosen global poles determined by the global structure of variety.
This method has already been applied extensively to four-dimensional
two-loop double-box integral and crossed-box integral, and to the
study of three-loop integrals and also integrals with doubled
propagators
\cite{Larsen:2012sx,CaronHuot:2012ab,Johansson:2012zv,Sogaard:2013yga,Johansson:2013sda,Sogaard:2013fpa,Sogaard:2014ila,Sogaard:2014oka}.

In both multi-loop integral reduction and integrand reduction
through an algebraic geometric approach, we can see that the equivalent
description of maximal unitarity cut, i.e., the simultaneous
solution (variety) of on-shell equations of propagators (ideal),
plays fundamental role. Although, in principle, such an
algebraic geometric approach can be applied to any loops,
the explicit application is
still limited to a few two-loop and three-loop diagrams due to the
complexity of computation. Thus, before a wider application to other
two-loop and three-loop diagrams, it would be better as an initial
step to study the global structure of varieties for all two-loop and
three-loop diagrams.

A four-dimensional $L$-loop amplitude has $4L$ degrees of freedom,
and it defines an integral in $\mathbb{C}^{4L}$ complex plane in the
algebraic geometry framework. By Hilbert's Nullstellensatz, the
number of propagators can be reduced to $n\leq 4L$. The polynomials
of~$n$~propagators for a given diagram define an ideal $I=\langle
f_1,f_2,\ldots,f_n\rangle$ in the polynomial ring
$\mathbb{C}[x_1,\ldots, x_{4L}]$. If $n=4L$, the ideal is
zero-dimensional, and the corresponding variety
is a finite set of points in $\mathbb{C}^{4L}$.
If $n=4L-1$, the ideal is one-dimensional, and the corresponding
variety is an algebraic curve. This curve may be reducible, and
could consist of several irreducible curves.
However, the algebraic curve can be
universally characterized by its geometric genus, which is a topological
invariant. For a specific diagram with $4L-1$ propagators, if the
algebraic curve defined by the variety has genus $k$, then the global
structure of variety is described by a $k$-fold torus or its
degenerate pictures. If $n<4L-1$, the ideal is higher-dimensional
and the corresponding higher-dimensional variety is more
complicated to study.

In \cite{Huang:2013kh}, the arithmetic genus and singular points of an
algebraic curve are introduced to study the geometric genus of
curves defined by one-loop, two-loop and some of three-loop
diagrams. In this paper, we generalize the study of global structure
to all four-dimensional three-loop diagrams with eleven propagators.
The Riemann-Hurwitz formula is applied to the study of genus, and an
algorithm based on numerical algebraic geometry~\cite{BertiniBook,Mehta:2012wk} is implemented to compute necessary terms in the
Riemann-Hurwitz formula. With this algorithm, it is possible to
study the global structure of curves defined by four-loop diagrams
efficiently. For some three-loop diagrams, a recursive formula
derived from the Riemann-Hurwitz formula is presented to study the genus
of three-loop diagrams recursively from the genus of two-loop diagrams,
where a lattice convex polytope method is adopted. As of theoretical
interests, some interesting phenomena regarding the genus of any loop orders
are explored. We hope that these results could be useful for the
integral and integrand reduction of three-loop amplitudes via
algebraic geometry in the near future.

The remainder of this paper is organized as follows.
In Section 2, we introduce the
Riemann-Hurwitz formula for the computation of geometric genus. An
algorithm based on numerical algebraic geometry is
also discussed for numerically computing the genus of any algebraic
curve. In Section 3, we re-study the global structure of curves of
two-loop diagrams by the Riemann-Hurwitz formula, and in Section 4, we
generalize the analysis to curves of certain three-loop diagrams
whose sub-two-loop diagram is double-box or crossed-box. A recursive
formula derived from the Riemann-Hurwitz formula is presented for
recursively computing genus of curve defined by three-loop diagrams
from genus of curve defined by two-loop diagrams. A proof of the
recursive formula is provided based on convex polytope
techniques. In Section 5, the genus of curves defined by the remaining
three-loop diagrams is analyzed by the algorithm. The genus of
curves defined by an infinite series of White-house diagrams to any loop
orders is also studied as an example of recursive formula for
higher loop diagrams. In Section 6, we summarize the
results and discuss
generalizations for future work.

%%%%%%%%%%%%%%%%%%%%
\section{Preliminary}
%%%%%%%%%%%%%%%%%%%%%%

%%%%%%%%%%%%%%%%%%
\subsection{The Riemann-Hurwitz formula and geometric genus}
%%%%%%%%%%%%%%%%%%%%%

The Riemann-Hurwitz formula describes the relation of the Euler
characteristic between two surfaces when one is a ramified covering
of the other. It is often applied to the theory of Riemann surfaces
and algebraic curves for finding the genus of a complicated Riemann
surface that maps to a simpler surface (for more mathematical
details, definition of geometric genus, properties of algebraic
curve and other relevant definitions see, e.g., the books
 \cite{hartshorne1977algebraic,maclean2007algebraic}).

The Euler characteristic $\chi$ is a topological invariant. For an
orientable surface, it is given by $\chi=2-2g$, where $g$ is the
genus. A covering map is a continuous function $f$ from a
topological space $S'$ to another topological space $S$
\bea f:~~S'\mapsto S~~~\nonumber\eea
such that each point in $S$ has an open neighborhood evenly covered
by $f$. In the case of an unramified covering map $f$ which is
surjective and of degree $\mbox{deg}[f]$, we have formula
\bea \chi_{S'}=\mbox{deg}[f]\cdot\chi_S~.~~~\eea
The ramification, roughly speaking, is the case when sheets come
together. The covering map $f$ is said to be ramified at point $P$
in $S'$ if there exist analytic coordinates near $P$ and $f(P)$ such
that $f$ takes the form $f(z)=z^n$, $n>1$. The number $n$ is the
ramification index $e_P$ at point $P$. The ramification of covering
map at some points introduces a correction to the above formula as
\bea \chi_{S'}=\mbox{deg}[f]\cdot \chi_S-\sum_{P\in
S'}(e_P-1)~,~~~\label{HurwitzFormula}\eea
known as the Riemann-Hurwitz formula. Applying this formula to the
case of algebraic curves, for a curve $\mathcal{C}'$ of genus
$g_{\mathcal{C}'}$ and another curve $\mathcal{C}$ of genus
$g_{\mathcal{C}}$, there is a (ramified) covering~map
\bea f:~~\mathcal{C}'\mapsto \mathcal{C}~,~~~\nonumber\eea
and the genus of two curves are related by
\bea
2g_{\mathcal{C}'}-2=\mbox{deg}[f](2g_{\mathcal{C}}-2)+\sum_{P\in
\mathcal{C}'}(e_P-1)~.~~~\label{HurwitzFormula2}\eea
Note that ramification can also happen at infinity.
Knowing the degree of covering map, the genus of curve $\mathcal{C}$
and the ramification points, it is possible to compute the genus of
curve $\mathcal{C}'$.

A special version of (\ref{HurwitzFormula2}) is that the covering
map $f$ maps a curve $\mathcal{C}$ to a curve of genus zero. In this
case, $\mbox{deg}[f]=\mbox{deg}[\mathcal{C}]$, and the
Riemann-Hurwitz formula (\ref{HurwitzFormula2}) is rewritten as
\bea g_{\mathcal{C}}=-\mbox{deg}[\mathcal{C}]+1+{1\over
2}\Big(\rho_{\infty}+\sum_{P\in
\mathcal{C}}\rho_{P}\Big)~,~~~\label{HurwitzFormula3}\eea
where $\rho_P=e_P-1$.

We can either use formula (\ref{HurwitzFormula2}) or formula
(\ref{HurwitzFormula3}) for the genus analysis. For an algebraic
curve $\mathcal{C}'$ defined by several polynomial
equations, simplifications can be made when a
subset of polynomial equations contain fewer variables
which also define a curve $\mathcal{C}$.
In this case, we can compute the degree of the
covering map $f: \mathcal{C}'\mapsto \mathcal{C}$ and the
ramification points. The computation is relatively simpler than the
case of mapping to a curve of genus zero, especially for the curves
defined by the maximal unitarity cut of multi-loop diagrams, where
analytic study is possible. However, it is always possible to
compute the genus of any algebraic systems of curves by formula
(\ref{HurwitzFormula3}), although the computation would become very
complicated. In the next subsection, we will describe an algorithm
for computing the geometric genus by formula
(\ref{HurwitzFormula3}), based on numerical algebraic geometry.

%%%%%%%%%%%%%%%%%%%%
\subsection{An algorithm for computing the geometric genus}
%%%%%%%%%%%%%%%%%%%%%%%
The algorithm for numerically computing the geometric genus of a
curve presented in  \cite{GeomGenus} (see also
 \cite[\S~2.6]{hsProjections} and
 \cite[\S~15.1,\S~16.5.2]{BertiniBook}) follows from the
Riemann-Hurwitz formula using numerical algebraic geometry
techniques to compute the necessary items in the formula. The
following provides a short description of the techniques needed to
describe this algorithm, namely witness sets, computing a superset
of the branchpoints, and monodromy, with the books
 \cite{BertiniBook,SW05} providing more details.

The input for the algorithm of  \cite{GeomGenus} to compute the
geometric genus of an irreducible curve $\mathcal{C}\subset\bC^n$ is
a witness set for $\mathcal{C}$. Let $f$ be a system of polynomials
in $n$ variables such that $\mathcal{C}$ is an irreducible component
of the set $\Var(f) = \{x~|~f(x) = 0\}$. A {\em witness set} for
$\mathcal{C}$ is the triple $\{f,\ell,W\}$ where $\ell$ is a general
linear polynomial and $W = \mathcal{C}\cap\Var(\ell)$.  The set $W$
is called a {\em witness point set} for $\mathcal{C}$ with $\deg
[\mathcal{C}] = |W|$. The concept of witness sets was described in
particle and string theory frameworks in
 \cite{Mehta:2011wj,Mehta:2012wk,Hauenstein:2012xs}.

By, for example, isosingular deflation  \cite{IsoSing}, we can
assume that $\mathcal{C}$ has multiplicity $1$ with respect to $f$.

Given a system $f$, we are interested in computing a witness set for
each curve $\mathcal{C}$ that is an irreducible component $\Var(f)$.
To accomplish this, we first select a general linear polynomial
$\ell$ and compute the set of isolated points $\sW$ in
$\Var(f,\ell)$, which is the union of the witness points sets
$W_{\mathcal{C}}$ for each such curve.  For example, one could use
regeneration~\cite{Regen,RegenCascade} and the local dimension test
 \cite{LDT} to yield such a set.

The set $\sW$ is partitioned into the various $W_{\mathcal{C}}$, for
example, using many random monodromy loops with the decomposition
confirmed using the trace test  \cite{Symmetric}. Since performing a
monodromy loop is a key aspect of computing the geometric genus, we
will summarize the computation here for curves. Let $\sH$ define a
loop of general hyperplanes in $\bC^n$ with $\sH(0) = \sH(1) =
\Var(\ell)$. Thus, $\Var(f)\cap\sH(t)$ defines a collection of
smooth paths $z(t)$ with $z(0),z(1)\in\sW$.  Since the points $z(0)$
and $z(1)$ must lie on the same irreducible component, this
monodromy loop provides information about how to partition $\sW$
when $z(0)\neq z(1)$.

Suppose that $\{f,\ell,W\}$ is a witness set for an irreducible
curve $\mathcal{C}$. Let $\pi:\bC^n\rightarrow\bC$ be a general
linear projection defined by $\pi(x) = \alpha\cdot x$ for
$\alpha\in\bC^n$. As shown in  \cite{GeomGenus}, a finite superset
of the branchpoints of $\mathcal{C}$ with respect to $\pi$ is
sufficient since the contribution in the Riemann-Hurwitz formula
from points which are not branchpoints is zero. In particular, such
a superset is the finite set of points $B_{\mathcal{C}}\subset
\mathcal{C}$ such that
$$\left[\begin{array}{c} Jf(x) \\ \alpha
\end{array}\right]$$ is rank deficient, where $J$ is the Jacobian matrix of the system $f(x)$. The set $B_{\mathcal{C}}$ can be
computed from a witness set for $\mathcal{C}$ using regeneration
extension  \cite{RegenExtend} with  \cite{CriticalPoints}.

For each distinct number in $\pi(B_{\mathcal{C}}) = \{\pi(b)~|~b\in
B_{\mathcal{C}}\}$, we need to compute the contribution
$\rho_{\pi(b)}$ for each $\pi(b)$ in the Riemann-Hurwitz formula.  A
monodromy loop surrounding $\pi(b)$ that does not include any other
point in $\pi(B_{\mathcal{C}})\setminus\pi(b)$ yields a
decomposition of the $\deg [\mathcal{C}]$ points into
$\gamma_{\pi(b)}$ sets with $\rho_{\pi(b)} = \deg [\mathcal{C}] -
\gamma_{\pi(b)}$. One also needs to perform a monodromy loop which
surrounds every point in $\pi(B)$ to compute the contribution
$\rho_{\infty}$ at $\infty$. Thus, the geometric genus of
$\mathcal{C}$ is
$$g_{\mathcal{C}} = -\deg [\mathcal{C}] + 1 + \frac{1}{2}\left(\rho_{\infty} + \sum_{\pi(b)\in\pi(B_{\mathcal{C}})} \rho_{\pi(b)}\right).$$

\begin{remark}
If one performs the algorithm described above for an input witness
set $\{f,\ell,W\}$ of a curve $\mathcal{C}$ which is not
irreducible, then the output is
$$g_{\mathcal{C}_1} + \cdots + g_{\mathcal{C}_k} - k + 1$$
where $\mathcal{C}_1,\dots,\mathcal{C}_k$ are the irreducible
components of $\mathcal{C}$. This value could be
negative.\footnote{We thank Andrew Sommese for communicating this
remark to us.}
\end{remark}

%%%%%%%%%%%%%%%%%%%%%%%%%%
\subsection{Curves of three-loop diagrams}
%%%%%%%%%%%%%%%%%%%%%%%%%%%%

In this subsection, we classify all the three-loop diagrams whose
maximal unitarity cuts yield an algebraic system that
defines a non-trivial curve. Naively, there are a large number of three-loop diagrams in
four-dimension, with the total number of propagators up to twelve.
Diagrams with more than twelve propagators are over-determined,
i.e., the number of propagators $n_{\ell_1\ell_2\ell_3}$ is larger
than the independent parametrization variables of loop momenta.
Thus,
they can be reduced to diagrams with twelve propagators or lower.
Similarly, the number of propagators containing only two of the loop
momenta $n_{\ell_1\ell_2}, n_{\ell_2\ell_3}$ or $n_{\ell_1\ell_3}$
should be smaller than eight, and the number of propagators
containing only one loop momentum $n_{\ell_1}, n_{\ell_2}$ or
$n_{\ell_3}$ should be smaller than four. If there are no shared
propagators between any two loops, i.e., different loops only be
connected at vertices, then the integral of three-loop diagrams can
be rewritten as product of an one-loop integral and a two-loop
integral, or product of three one-loop integrals. So, the topology
defined by these diagrams is the same as the one defined by two-loop
diagrams or one-loop diagrams. We are only interested in the
non-trivial three-loop diagrams with loops being connected by shared
propagators.

\begin{figure}
\centering
  % Requires \usepackage{graphicx}
  \includegraphics[width=5.5in]{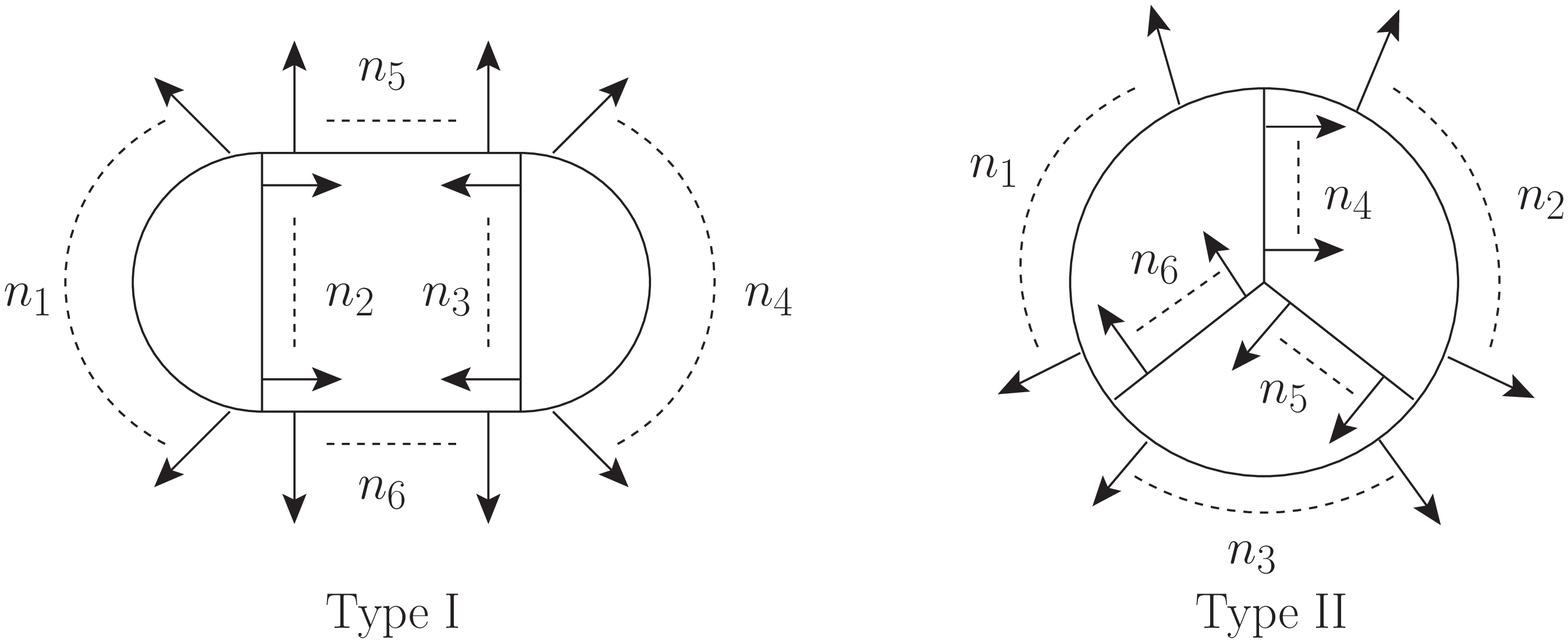}\\
  \caption{Basic topologies of three-loop ladder type diagrams and Mercedes-logo type
  diagrams. $n_i$ is the number of propagators along the dashed line. All external momenta
  are massive, and all vertices are attached by external legs, which are not explicitly drawn
  in the figure.}\label{3LoopTopo}
\end{figure}

Basically, there are two types of non-trivial three-loop diagrams as
shown in Figure (\ref{3LoopTopo}). Type I diagram is the ladder type
diagram, where loops $(\ell_1,\ell_2)$, $(\ell_2,\ell_3)$ have
shared propagators, while $(\ell_1,\ell_3)$ do not have shared
propagators. Type II diagram is the Mercedes-logo type diagram,
where any two loops have shared propagators. These diagrams could be
planar or non-planar diagrams, according to the value of $n_i$,
where $n_i$ is the number of propagators along the dashed lines in
Figure (\ref{3LoopTopo}). In the current paper, we are interested in
the topologies whose maximal unitarity cuts define a curve. So we
require the number of propagators to be $\sum_{i=1}^6n_i=11$.

For type I diagram, we take the convention that the left loop is
$\ell_1$, the middle loop is~$\ell_2$ and the right loop is
$\ell_3$. Then we have the following inequalities for $n_i$,
\bea &&n_{\ell_1\ell_2}=n_1+n_2+n_5+n_6\leq
8~~,~~n_{\ell_2\ell_3}=n_3+n_4+n_5+n_6\leq 8~.~~~\nonumber\eea
Of course we have assumed every $1\leq n_i\leq 4$ except $0\leq
n_5\leq 4$, in order to generate all ladder type diagrams. However,
due to the symmetries of diagrams, there will be over-counting from
the solution of above inequalities. In order to remove the
over-counting, we further require that
\bea n_1\geq n_2~~,~~n_4\geq n_3~~,~~n_6\geq n_5~.~~~\eea
These inequalities remove the over-counting from symmetries inside
each left loop, middle loop and right loop. However, there is still
symmetry between the left and right loops. The over-counting of this
symmetry can be removed by following two sets of inequalities
\bea \mbox{(1)}~~n_1=n_4~~,~~n_2\geq
n_3~~,~~~\mbox{(2)}~~n_1>n_4~.~~~\eea
The above inequalities generate 36 diagrams, denoted by
$(n_1,n_2,n_3,n_4,n_5,n_6)$ as
\bea
&&(2,2,1,2,0,4)~,~(2,2,1,2,1,3)~,~(2,2,1,2,2,2)~,~(2,2,2,2,0,3)~,~(2,2,2,2,1,2)~,~~~\nonumber\\
&&(3,1,1,2,0,4)~,~(3,1,1,2,1,3)~,~(3,1,1,2,2,2)~,~(3,1,1,3,0,3)~,~(3,1,1,3,1,2)~,~~~\nonumber\\
&&(3,1,2,2,0,3)~,~(3,1,2,2,1,2)~,~(3,2,1,2,0,3)~,~(3,2,1,2,1,2)~,~(3,2,1,3,0,2)~,~~~\nonumber\\
&&(3,2,1,3,1,1)~,~(3,2,2,2,0,2)~,~(3,2,2,2,1,1)~,~(3,2,2,3,0,1)~,~(3,3,1,2,0,2)~,~~~\nonumber\\
&&(3,3,1,2,1,1)~,~(3,3,1,3,0,1)~,~(3,3,2,2,0,1)~,~(4,1,1,2,0,3)~,~(4,1,1,2,1,2)~,~~~\nonumber\\
&&(4,1,1,3,0,2)~,~(4,1,1,3,1,1)~,~(4,1,1,4,0,1)~,~(4,1,2,2,0,2)~,~(4,1,2,2,1,1)~,~~~\nonumber\\
&&(4,1,2,3,0,1)~,~(4,2,1,2,0,2)~,~(4,2,1,2,1,1)~,~(4,2,1,3,0,1)~,~(4,2,2,2,0,1)~,~~~\nonumber\\
&&(4,3,1,2,0,1)~.~~~\nonumber\eea
However, if any $n_{\ell_i}=4$, then the corresponding loop momentum
$\ell_i$ can be completed determined by the equations of unitarity
cuts. So this loop momentum is effectively the external momentum for
the remaining loops. In this case, the curve associated with
three-loop diagram is reduced to the curve associated with two-loop
diagram. Similarly, if any $n_{\ell_i\ell_j}=8$, then the
corresponding loop momenta $\ell_i,\ell_j$ can be completely
determined. The curve is reduced to the one associated with one-loop
diagram. Among the 36 diagrams, there are still 13 diagrams whose
curves can not be reduced to the ones associated with one-loop or
two-loop diagrams. We shall study the topologies of these diagrams
in the following sections.

For type II diagram, we take the convention that the left-top loop
is $\ell_1$, the right-top loop is $\ell_2$ and the bottom loop is
$\ell_3$. Again we have
\bea &&n_{\ell_1\ell_2}=n_1+n_2+n_4\leq
8~~,~~n_{\ell_2\ell_3}=n_2+n_3+n_5\leq
8~~,~~n_{\ell_1\ell_3}=n_1+n_3+n_6\leq 8~.~~~\nonumber\eea
Also we have $1\leq n_i\leq 4$. This type of diagrams has symmetries
by exchanging $(n_1\leftrightarrow n_4, n_3\leftrightarrow n_5)$, or
$(n_2\leftrightarrow n_4, n_3\leftrightarrow n_6)$ or
$(n_1\leftrightarrow n_6, n_2\leftrightarrow n_5)$. By considering
these symmetries, we can generate 15 diagrams, denoted by
$(n_1,n_2,n_3,n_4,n_5,n_6)$ as
\bea
&&(2,2,3,1,1,2)~,~(2,2,3,2,1,1)~,~(2,1,3,2,1,2)~,~(2,1,3,2,2,1)~,~(2,2,2,2,2,1)~,~~~\nonumber\\
&&(3,2,3,1,1,1)~,~(3,1,3,2,1,1)~,~(3,1,3,1,1,2)~,~(1,2,3,3,1,1)~,~(3,1,4,1,1,1)~,~~~\nonumber\\
&&(1,1,4,3,1,1)~,~(2,2,4,1,1,1)~,~(2,1,4,1,2,1)~,~(2,1,4,1,1,2)~,~(2,1,4,2,1,1)~.~~~\nonumber\eea
For diagrams with $n_{\ell_i\ell_j}=8$ or $n_{\ell_i}=4$, the curves
are reduced to the ones associated with one-loop triangle, two-loop
double-box or crossed-box diagrams. Among the 15 diagrams, there are
eight diagrams which can not be reduced. These are the eight
three-loop Mercedes-logo diagrams which we will study in the
following sections.

In summary, there are in total $13+8=21$ three-loop diagrams
generating algebraic systems defining non-trivial curves. Among them, 16
diagrams have a sub-two-loop diagram whose maximal unitarity cuts
also define curves. For these diagrams, we will present a recursive
formula based on Riemann-Hurwitz formula, to compute the genus
recursively from two-loop diagrams. The remaining five diagrams can
not be computed by the recursive formula, so we will use the
algorithm based on numerical algebraic geometry to study~the~genus.

%%%%%%%%%%%%%%%%%%%%%
\section{Counting the ramified points of two-loop diagrams}
%%%%%%%%%%%%%%%%%%%%%

As a warm-up exercise for three-loop analysis, let us briefly go
through the study of two-loop diagrams in the framework of
Riemann-Hurwitz formula (\ref{HurwitzFormula2}). There are two
diagrams whose equations of maximal unitarity cuts define
non-trivial irreducible curves. As it is well studied in the
literature
\cite{Kosower:2011ty,CaronHuot:2012ab,Feng:2012bm,Sogaard:2013yga},
the curve associated with the double-box diagram has genus one and
the curve associated with the crossed-box diagram has genus three,
obtained by directly computing the arithmetic genus and singular
points of the curves, or inferred from the picture of Riemann
spheres in the limit of  degenerate kinematics.

\begin{figure}
  % Requires \usepackage{graphicx}
  \centering
  \includegraphics[width=4in]{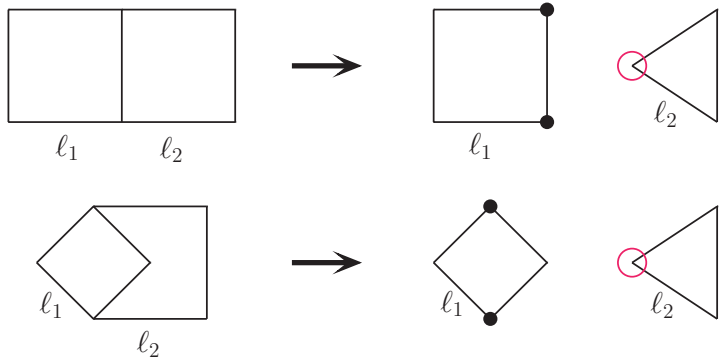}\\
  \caption{Two-loop double-box diagram and crossed-box diagram constructed from
  one-loop box and triangle diagrams by connecting them in two different ways.}\label{construct2L}
\end{figure}

Notice that these two diagrams can be constructed from a box diagram
and a triangle diagram as shown in Figure (\ref{construct2L}), by
opening the vertex in the triangle diagram marked as red circle and
connecting the two legs to the box diagram at the vertices marked as
black dots respectively. If ignoring all equations from the box
diagram, the equation system associated with a triangle diagram itself
defines a curve. Referring to the Riemann-Hurwitz formula, the genus
of the curve associated with the double-box diagram or crossed-box
diagram is related to the genus of curve associated with triangle
diagram, by considering the~covering~map
\bea f:~~
\mathcal{C}_{\includegraphics[width=1.3em]{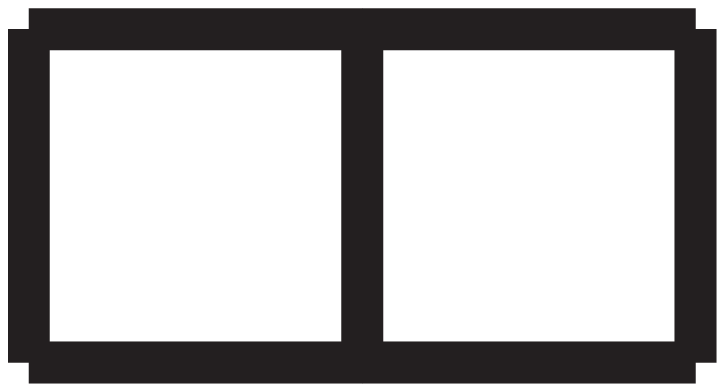}}\mapsto
\mathcal{C}_{\triangle}~~~~\mbox{or}~~~~f:~~
\mathcal{C}_{\includegraphics[width=1em]{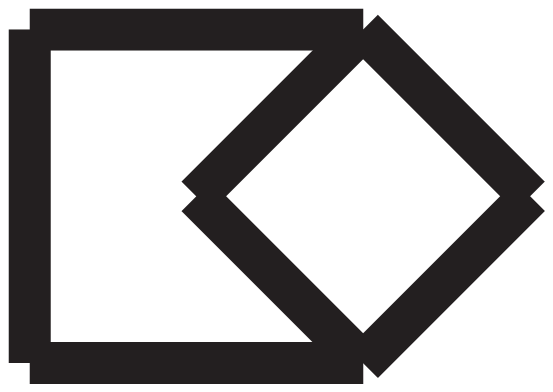}}\mapsto
\mathcal{C}_{\triangle}~.~~~\label{2LoopMap1Loop}\eea
The cut equations of triangle diagram are given by
\bea
\ell^2=0~~,~~(\ell-K_1)^2-\ell^2=0~~,~~(\ell-K_1-K_2)^2-\ell^2=0~.~~~\nonumber\eea
The latter two equations are linear in $\ell$, so there is only one
quadratic equation after some algebraic manipulation of above three
equations. By solving two variables with two linear equations, the
remaining quadratic equation becomes equation of conics, and it is
topological equivalent to genus zero Riemann sphere. So, the only
lacking data for computing the genus of double-box or crossed-box
diagram is the ramified points of covering map
(\ref{2LoopMap1Loop}). Since these two-loop diagrams have been
separated into two parts $\mathcal{C}_{\triangle}$ and
$\mathcal{P}_{\square}$, for any given point $P_i$ in the curve
$\mathcal{C}_{\triangle}$, the four equations of
$\mathcal{P}_{\square}$ define points
$\{P_{i,1},P_{i,2},\ldots,P_{i,m}\}$ in
$\mathcal{C}_{\includegraphics[width=1.3em]{dboxSym.eps}}$ or
$\mathcal{C}_{\includegraphics[width=1em]{cboxSym.eps}}$. This means
that the covering map (\ref{2LoopMap1Loop}) maps
\bea \{P_{i,1},P_{i,2},\ldots,P_{i,m}\}\mapsto P_i~~~~\nonumber \eea
from curves of two-loop diagrams to curve of triangle diagram. If
all $\{P_{i,1},P_{i,2},\ldots,P_{i,m}\}$ are the same point, then
the map $f$ is ramified at the point $P_i$, and the ramification
index is $m$.

We need to compute the ramified points and their ramification
indices in the covering map (\ref{2LoopMap1Loop}). We use the same
parametrization of loop momenta as in \cite{Feng:2012bm}, and define
$\mathbf{x}=\{x_1,x_2,x_3,x_4\}$ and
$\mathbf{y}=\{y_1,y_2,y_3,y_4\}$ as the parametrization variables of
$\ell_1$ and $\ell_2$ respectively. A function
$h(\mathbf{a}^{m_1}\mathbf{b}^{m_2})$ with argument
$\mathbf{a}^{m_1}\mathbf{b}^{m_2}$ denotes a function whose highest
degree monomials are terms of
$\prod_{k=1}^{m_1}a_{i_k}\prod_{k=1}^{m_2}b_{j_k}$, where $a_{i_k},
b_{i_k}$ could be any elements in $\mathbf{a},\mathbf{b}$. So
$h(\mathbf{x}^2)$ is a quadratic function of $x_i$, while
$h(\mathbf{x}\mathbf{y})$ is also a quadratic function of $x_iy_j$,
but a linear function with respect to $x_i$ or $y_i$ individually.
Among the seven equations of maximal unitarity cuts, there are four
linear equations and three quadratic equations. Using the linear
equations, we can always solve two of $x_i$'s and two of $y_i$'s.
Define the remaining variables as $\mathbf{x}_s=\{x_1,x_2\},
\mathbf{y}_s=\{y_1,y_2\}$, then the remaining three quadratic
equations are equations of $\mathbf{x}_s$ and $\mathbf{y}_s$. The
covering map (\ref{2LoopMap1Loop}) actually maps
\bea \mathcal{C}_{\includegraphics[width=1.3em]{dboxSym.eps}}:
\left\{\begin{array}{l}
                                                                 Q_1(\mathbf{x}^2)=0 \\
                                                                 L_1(\mathbf{x})=0 \\
                                                                 L_2(\mathbf{x})=0 \\
                                                                 Q_2(\mathbf{x}\mathbf{y})=0 \\
                                                                 Q_3(\mathbf{y}^2)=0 \\
                                                                 L_3(\mathbf{y})=0 \\
                                                                 L_4(\mathbf{y})=0
                                                               \end{array}\right.~~~~\mbox{or}~~~~
                                                               \mathcal{C}_{\includegraphics[width=1em]{cboxSym.eps}}:
\left\{\begin{array}{l}
                                                                 Q_1(\mathbf{x}^2)=0 \\
                                                                 L_1(\mathbf{x})=0 \\
                                                                 Q_2(\mathbf{x}\mathbf{y})=0 \\
                                                                 L_2(\mathbf{x},\mathbf{y})=0 \\
                                                                 Q_3(\mathbf{y}^2)=0 \\
                                                                 L_3(\mathbf{y})=0 \\
                                                                 L_4(\mathbf{y})=0
                                                               \end{array}\right.\mapsto
                                                               \mathcal{C}_{\triangle}:
                                                               \left\{\begin{array}{l}
                                                               Q_3(\mathbf{y}^2)=0 \\
                                                               L_3(\mathbf{y})=0 \\
                                                               L_4(\mathbf{y})=0
                                                               \end{array}\right.~.~~~
                                                               \eea
Solving the linear equations in double-box or crossed-box diagram,
we get
\bea \mathbf{x}\mapsto \mathbf{x}_s~~,~~\mathbf{y}\mapsto
\mathbf{y}_s~~\mbox{or}~~\mathbf{x}\mapsto
\mathbf{x}_s,\mathbf{y}_s~~,~~\mathbf{y}\mapsto
\mathbf{y}_s~.~~~\nonumber\eea
So we can simplify the covering map as
\bea \mathcal{C}_{\includegraphics[width=1.3em]{dboxSym.eps}}:
\left\{\begin{array}{l}
                                                                 Q_1(\mathbf{x}_s^2)=0 \\
                                                                 Q_2(\mathbf{x}_s\mathbf{y}_s)=0 \\
                                                                 Q_3(\mathbf{y}_s^2)=0
                                                                 \end{array}\right.~~~~\mbox{or}~~~~
                                                               \mathcal{C}_{\includegraphics[width=1em]{cboxSym.eps}}:
\left\{\begin{array}{l}
                                                                 Q_1(\mathbf{x}_s^2,\mathbf{x}_s\mathbf{y}_s,\mathbf{y}_s^2)=0 \\
                                                                 Q_2(\mathbf{x}_s\mathbf{y}_s,\mathbf{y}_s^2)=0 \\
                                                                 Q_3(\mathbf{y}_s^2)=0
                                                               \end{array}\right.\mapsto
                                                               \mathcal{C}_{\triangle}:
                                                               \begin{array}{l}
                                                               Q_3(\mathbf{y}_s^2)=0
                                                               \end{array}~.~~~
                                                               \eea
Since $Q_1$ is quadratic in $\mathbf{x}_s$ but $Q_2$ is linear in
$\mathbf{x}_s$, they define two covering sheets over Riemann sphere
$\mathcal{C}_{\triangle}$, so the covering map is a double cover.
For any given point $P_i=\{y^P_1,y^P_2\}$ in the curve
$\mathcal{C}_{\triangle}$, the joint equations $Q_1=Q_2=0$ can be
used to solve $\{x_1,x_2\}$, and it has two solutions because of its
quadratic property. Generally the two solutions are distinct,
however when the discriminant equals to zero, they coincide in the
same point and produce a ramified point with ramification index
$e_P=2$.

Let us generically consider two equations
\bea
a_1x_1^2+a_2x_1x_2+a_3x_2^2+a_4x_1+a_5x_2+a_0=0~~,~~b_1x_1+b_2x_2+b_0=0~.~~~\nonumber\eea
The discriminant $\Delta$ is
\bea \Delta&=&a_2^2 b_0^2-4 a_1 a_3 b_0^2+4 a_3 a_4 b_0 b_1-2 a_2
a_5 b_0 b_1-4 a_0 a_3 b_1^2+a_5^2 b_1^2\nonumber\\
&&-2 a_2 a_4 b_0 b_2+4 a_1 a_5 b_0 b_2+4 a_0 a_2 b_1 b_2-2 a_4 a_5
b_1 b_2-4 a_0 a_1 b_2^2+a_4^2 b_2^2~.~~~\label{2loopDelta}\eea
A given point $P_i$ in curve $\mathcal{C}_{\triangle}:
Q_3(\mathbf{y}_s^2)=0$ should also follow the constraint
$\Delta(y_1,y_2)=0$, if it is a ramified point. So these two
equations completely determine the location of ramified points. In
the double-box case, all $a_i$'s are independent of $\mathbf{y}_s$,
while $b_i$'s are linear functions of $\mathbf{y}_s$, so the
discriminant is a generic quadratic function of $\mathbf{y}_s$. By
B\'ezout's theorem, the two equations define $2\times 2=4$ distinct
points, which are the ramified points with index $e_P=2$. In the
crossed-box case, $a_1,a_2,a_3$ are independent of $\mathbf{y}_s$,
$a_4,a_5,b_1,b_2$ are linear in $\mathbf{y}_s$ and $a_0,b_0$ are
quadratic in $\mathbf{y}_s$, so $\Delta(y_1,y_2)$ is a generic
function of degree four in $\mathbf{y}_s$. These two equations
define $2\times 4=8$ ramified points with ramification index
$e_P=2$. Using Riemann-Hurwitz formula, we get
\bea
&&2g_{\includegraphics[width=1.3em]{dboxSym.eps}}-2=2(2g_{\triangle}-2)+4(2-1)~~\to~~g_{\includegraphics[width=1.3em]{dboxSym.eps}}=1~,~~~\nonumber\\
&&2g_{\includegraphics[width=1em]{cboxSym.eps}}-2=2(2g_{\triangle}-2)+8(2-1)~~\to~~g_{\includegraphics[width=1em]{cboxSym.eps}}=3~,~~~\nonumber\eea
which agree with the known results in the literature
\cite{Kosower:2011ty,CaronHuot:2012ab,Feng:2012bm,Sogaard:2013yga}.

To summarize, in order to compute the genus of curve associated with
two-loop diagrams from genus of curve associated with one-loop
diagram, we separate the equations of maximal unitarity cuts into
$\mathcal{P}_{\square}$ and $\mathcal{C}_{\triangle}$. For given
point in $\mathcal{C}_{\triangle}$, equations of
$\mathcal{P}_{\square}$ always give two distinct solutions unless
the discriminant of $\mathcal{P}_{\square}$ is zero. This additional
constraint together with curve equations $\mathcal{C}_{\triangle}$
provide all information about the ramified points.

%%%%%%%%%%%%%%%%%%%%%
\section{Counting the ramified points of three-loop diagrams}
%%%%%%%%%%%%%%%%%%%%%
The same discussion can be generalized to compute the genus of
curves associated with three-loop diagrams from genus of curves
associated with two-loop diagrams, if the three-loop diagram has a
sub-two-loop which also defines a curve. For these three-loop
diagrams, we can always separate cut equations into
$\mathcal{P}_{\square}$ together with
$\mathcal{C}_{\includegraphics[width=1.3em]{dboxSym.eps}}$ or
$\mathcal{C}_{\includegraphics[width=1em]{cboxSym.eps}}$. Since
$g_{\includegraphics[width=1.3em]{dboxSym.eps}}$ and
$g_{\includegraphics[width=1em]{cboxSym.eps}}$ are known, the only
data we need to know is the ramified points. For ladder type
diagrams, among the eleven cut equations, there are five quadratic
equations and six linear equations, while for Mercedes-logo type
diagrams, there are six quadratic equations and five linear
equations.

We will discuss how to count the ramified points for these two types
of diagrams in this section. Defining
$\mathbf{x}=\{x_1,x_2,x_3,x_4\}$, $\mathbf{y}=\{y_1,y_2,y_3,y_4\}$
and $\mathbf{z}=\{z_1,z_2,z_3,z_4\}$ as parametrization variables
for $\ell_1,\ell_2,\ell_3$ respectively, where $\ell_1$ is the loop
momentum in box diagram, the number of ramified points is given by
\bea N&=&{8u\big(1-m_{xz}+(m_{xy}+m_{xz})(1-uv)\big)(1+m'_y)}\label{Ramified}\\
&&+{8uv(m_{xy}+m_{xz})(1+m'_{yz})}+{8v\big(1-m_{xy}+(m_{xy}+m_{xz})(1-uv)\big)(1+m'_z)}~.~~~\nonumber\eea
where $u=n_{xy}-m_{xy}, v=n_{nz}-m_{xz}$, and the ramification
indices are $e_P=2$. $n_{x}$, $n_{y}$, $n_{z}$, $n_{xy}$, $n_{xz}$,
$n_{yz}$ are the number of equations containing $\{\mathbf{x}\}$,
$\{\mathbf{y}\}$, $\{\mathbf{z}\}$, $\{\mathbf{x},\mathbf{y}\}$,
$\{\mathbf{x},\mathbf{z}\}$, $\{\mathbf{y},\mathbf{z}\}$
respectively, and $m_{x}$, $m_{y}$, $m_{z}$, $m_{xy}$, $m_{xz}$,
$m_{yz}$ are the number of linear equations containing
$\{\mathbf{x}\}$, $\{\mathbf{y}\}$, $\{\mathbf{z}\}$,
$\{\mathbf{x},\mathbf{y}\}$, $\{\mathbf{x},\mathbf{z}\}$,
$\{\mathbf{y},\mathbf{z}\}$ respectively. Also
\bea m'_{y}=\Big\lfloor{3-m_{y}\over
2}\Big\rfloor~~,~~m'_{z}=\Big\lfloor{3-m_{z}\over
2}\Big\rfloor~~,~~m'_{yz}=\Big\lfloor{3-m_{yz}\over
2}\Big\rfloor~,~~~\nonumber\eea
where $\lfloor a\rfloor$ is the floor function giving the integer
part of $a$.

%%%%%%%%%%%%%%%%%
\subsection{Ladder type diagrams}
\begin{figure}
  % Requires \usepackage{graphicx}
  \centering
  \includegraphics[width=6in]{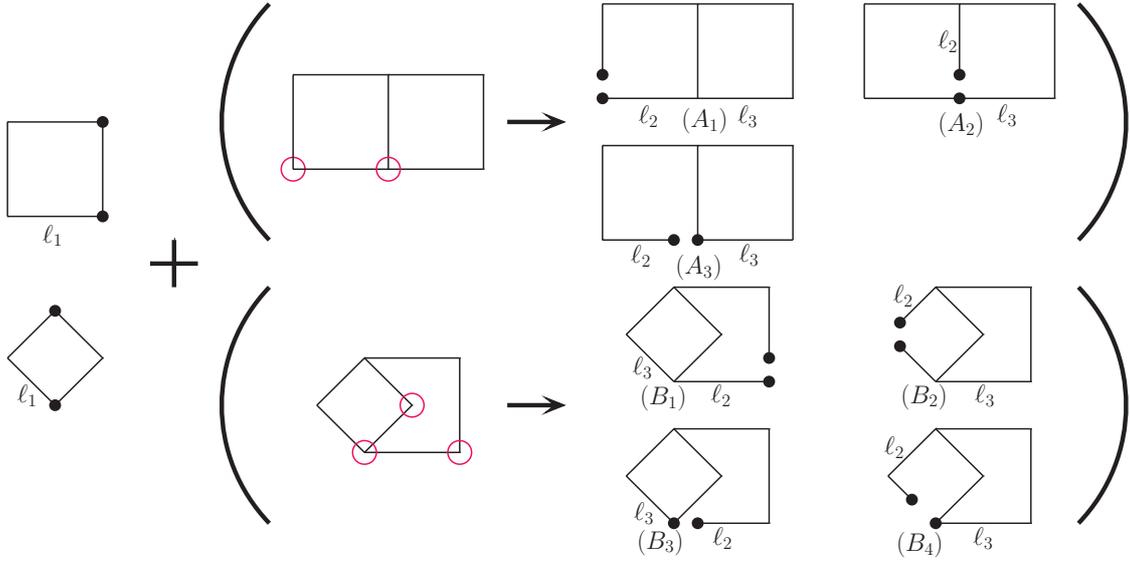}\\
  \caption{Different ways of connecting one-loop box diagram and two-loop double-box, crossed-box
  diagrams to construct three-loop ladder type diagrams. Vertices of two-loop diagrams marked as red circles are opened,
   and the corresponding internal lines are connected to the one-loop box diagram at the vertices marked as black dots.}\label{construct3LT1}
\end{figure}

%%%%%%%%%%%%%%%%%
The ladder type diagrams can be constructed from inserting box
diagram into two-loop diagrams at the vertices marked as red circles
as shown in Figure (\ref{construct3LT1}). Depending on the way of
opening the vertices, there are in total seven different ways
connecting to the two-loop diagrams, marked as black dots in the
seven diagrams in Figure (\ref{construct3LT1}).
For this type diagrams, we take the convention that $\ell_1,\ell_2$
have shared propagators, thus $u=1,v=0$. Then the formula
(\ref{Ramified}) is simplified to
\bea N&=&{8(1+m_{xy})(1+m'_{y})}~.~~~\label{ladderR}\eea
Since $m_{xy}, m'_{y}$ can either be one or zero, from
(\ref{ladderR}) we see that $N$ could be $8, 16$ and $32$. It is
also interesting to notice that $N$ picks up no information in the
$\ell_3$ loop. In fact, since $(1+m)=0$ if $m=0$ and $(1+m)=2$ if
$m=1$, we can artificially write $(1+m)=2^{m}$. Then (\ref{ladderR})
can be expressed as
\bea N&=&8\times 2^{m_{xy}}\times
2^{m'_{y}}~.~~~\label{ladderR2}\eea
This reformulation provides a diagrammatic meaning for the counting
of ramified points which we will show below.

For the ladder type diagrams, the covering map from curves
associated with three-loop diagrams to curves associated with
two-loop diagrams is given by
\bea
\mathcal{C}_{\square,\Diamond+\includegraphics[width=1.3em]{dboxSym.eps},\includegraphics[width=1em]{cboxSym.eps}}:~\left\{\begin{array}{l}
       Q_1(\mathbf{x}^2)=0\\
       Q_2(\mathbf{x}\mathbf{y})=0\\
       L'_1=0\\
       L'_2=0\\
       Q_3(\mathbf{y}^2)=0\\
       Q_4(\mathbf{y}\mathbf{z})=0\\
       Q_5(\mathbf{z}^2)=0\\
       L_1=0\\
       L_2=0\\
       L_3=0\\
       L_4=0
     \end{array}\right. \mapsto \mathcal{C}_{\includegraphics[width=1.3em]{dboxSym.eps},\includegraphics[width=1em]{cboxSym.eps}}:~\left\{\begin{array}{l}
       Q_3(\mathbf{y}^2)=0\\
       Q_4(\mathbf{y}\mathbf{z})=0\\
       Q_5(\mathbf{z}^2)=0\\
       L_1=0\\
       L_2=0\\
       L_3=0\\
       L_4=0
     \end{array}\right.~.~~~\eea
Although $Q_2(\mathbf{x}\mathbf{y})=0$ is a quadratic equation, it
is linear in $\mathbf{x}$, so there is only one quadratic equation
in $\mathbf{x}$. Anyway, equations
$Q_1(\mathbf{x}^2)=Q_2(\mathbf{x}\mathbf{y})=L'_1=L'_2=0$ define two
covering sheets over Riemann surface
$\mathcal{C}_{\includegraphics[width=1.3em]{dboxSym.eps}}$ or
$\mathcal{C}_{\includegraphics[width=1em]{cboxSym.eps}}$, just as in
the analysis of mapping two-loop diagrams to one-loop diagrams. So
it is a double cover. Points in the curve defined by
$Q_3=Q_4=Q_5=0$, $L_i=0, i=1,2,3,4$ become ramified points if they
follow the additional constraint $\Delta(y_1,y_2,z_1,z_2)=0$, which
is the discriminant (\ref{2loopDelta}) of $Q_1=Q_2=L'_1=L'_2=0$.

Equations $\Delta=Q_3=Q_4=Q_5=0$, $L_i=0,i=1,2,3,4$ define a
zero-dimensional ideal $I=\langle
\Delta,Q_3,Q_4,Q_5,L_1,L_2,L_3,L_4\rangle$ in polynomial ring
$\mathbb{C}[y_1,y_2,y_3,y_4,z_1,z_2,z_3,z_4]$, and the number of
distinct solutions equals to the degree of ideal. The up-bound of
distinct point solutions is
$\mbox{deg}[\Delta]\mbox{deg}[Q_3]\mbox{deg}[Q_4]\mbox{deg}[Q_5]$.
Numerically, the degree of ideal can be computed by the Gr\"{o}bner
basis of ideal, which is the degree of leading term in Gr\"{o}bner
basis, by many algorithms (e.g., using {\tt Macaulay2} \cite{M2}).
However, we want to compute the ramified points without explicit
computations. Notice that among the seven cut equations of
sub-two-loop part, only the four linear equations are different. The
linear equations of seven diagrams in Figure (\ref{construct3LT1})
are given by
\bea \left\{\begin{array}{l}
       L^{A_1}_1(\mathbf{y})=0 \\
       L^{A_1}_2(\mathbf{y})=0 \\
       L^{A_1}_3(\mathbf{z})=0 \\
       L^{A_1}_4(\mathbf{z})=0
     \end{array}\right.~~,~~\left\{\begin{array}{l}
       L^{A_2}_1(\mathbf{y},\mathbf{z})=0 \\
       L^{A_2}_2(\mathbf{y},\mathbf{z})=0 \\
       L^{A_2}_3(\mathbf{z})=0 \\
       L^{A_2}_4(\mathbf{z})=0
     \end{array}\right.~~,~~\left\{\begin{array}{l}
       L^{A_3}_1(\mathbf{y})=0 \\
       L^{A_3}_2(\mathbf{y})=0 \\
       L^{A_3}_3(\mathbf{z})=0 \\
       L^{A_3}_4(\mathbf{z})=0
     \end{array}\right.~,~~~\nonumber\eea
and
\bea \left\{\begin{array}{l}
       L^{B_1}_1(\mathbf{y})=0 \\
       L^{B_1}_2(\mathbf{y})=0 \\
       L^{B_1}_3(\mathbf{z})=0 \\
       L^{B_1}_4(\mathbf{y},\mathbf{z})=0
     \end{array}\right.~~,~~\left\{\begin{array}{l}
       L^{B_2}_1(\mathbf{y})=0 \\
       L^{B_2}_2(\mathbf{y},\mathbf{z})=0 \\
       L^{B_2}_3(\mathbf{z})=0 \\
       L^{B_2}_4(\mathbf{z})=0
     \end{array}\right.~~,~~\left\{\begin{array}{l}
       L^{B_3}_1(\mathbf{y})=0 \\
       L^{B_3}_2(\mathbf{y})=0 \\
       L^{B_3}_3(\mathbf{z})=0 \\
       L^{B_3}_4(\mathbf{y},\mathbf{z})=0
     \end{array}\right.~~,~~\left\{\begin{array}{l}
       L^{B_4}_1(\mathbf{y})=0 \\
       L^{B_4}_2(\mathbf{y},\mathbf{z})=0 \\
       L^{B_4}_3(\mathbf{z})=0 \\
       L^{B_4}_4(\mathbf{z})=0
     \end{array}\right.~.~~~\nonumber\eea
It is clear that for $A_1,A_3,B_1,B_3$, $m'_y=0$ and for
$A_2,B_2,B_4$, $m'_y=1$. We can assign a factor
\bea N_{\ominus}=1~~\mbox{to}~~A_1, A_3, B_1,
B_3~~,~~N_{\ominus}=2~~\mbox{to}~~A_2, B_2, B_4~,~~~\eea
in Figure (\ref{construct3LT1}). For the box diagram part, we have
$m_{xy}=0$ for $\mathcal{P}_{\square}$ and $m_{xy}=1$ for
$\mathcal{P}_{\Diamond}$. So we can assign a factor
\bea
N_{\bigcirc}=1~~\mbox{to}~~P_{\square}~~\mbox{and}~~N_{\bigcirc}=2~~\mbox{to}~~P_{\Diamond}~.~~~\eea

The genus of curves associated with these three-loop ladder type
diagrams can be computed from genus of curves associated with
two-loop double-box or crossed-box diagram via Riemann-Hurwitz
formula as \iffalse % \bea
2g_{\square,\Diamond+\includegraphics[width=1.3em]{dboxSym.eps},
\includegraphics[width=1em]{cboxSym.eps}}-2=2(2g_{\includegraphics[width=1.3em]{dboxSym.eps},
\includegraphics[width=1em]{cboxSym.eps}}-2)+8(1+m_{xy})(1+m'_{y})~,~~~\label{3to2recur1}\eea
%
\fi
%
\bea g_{\square,\Diamond+\includegraphics[width=1.3em]{dboxSym.eps},
\includegraphics[width=1em]{cboxSym.eps}}=2g_{\includegraphics[width=1.3em]{dboxSym.eps},
\includegraphics[width=1em]{cboxSym.eps}}-1+4(1+m_{xy})(1+m'_{y})~,~~~\label{3to2recur1}\eea
or diagrammatically as \iffalse % \bea
2g_{\square,\Diamond+\includegraphics[width=1.3em]{dboxSym.eps},
\includegraphics[width=1em]{cboxSym.eps}}-2=2(2g_{\includegraphics[width=1.3em]{dboxSym.eps},
\includegraphics[width=1em]{cboxSym.eps}}-2)+8N_{\bigcirc}\times
N_{\ominus}~.~~~\label{3to2recur2}\eea
%
\fi
%
\bea g_{\square,\Diamond+\includegraphics[width=1.3em]{dboxSym.eps},
\includegraphics[width=1em]{cboxSym.eps}}=2g_{\includegraphics[width=1.3em]{dboxSym.eps},
\includegraphics[width=1em]{cboxSym.eps}}-1+4N_{\bigcirc}\times
N_{\ominus}~.~~~\label{3to2recur2}\eea
The computation can be done by just looking at the diagrams.

\begin{figure}
  % Requires \usepackage{graphicx}
  \centering
  \includegraphics[width=5.5in]{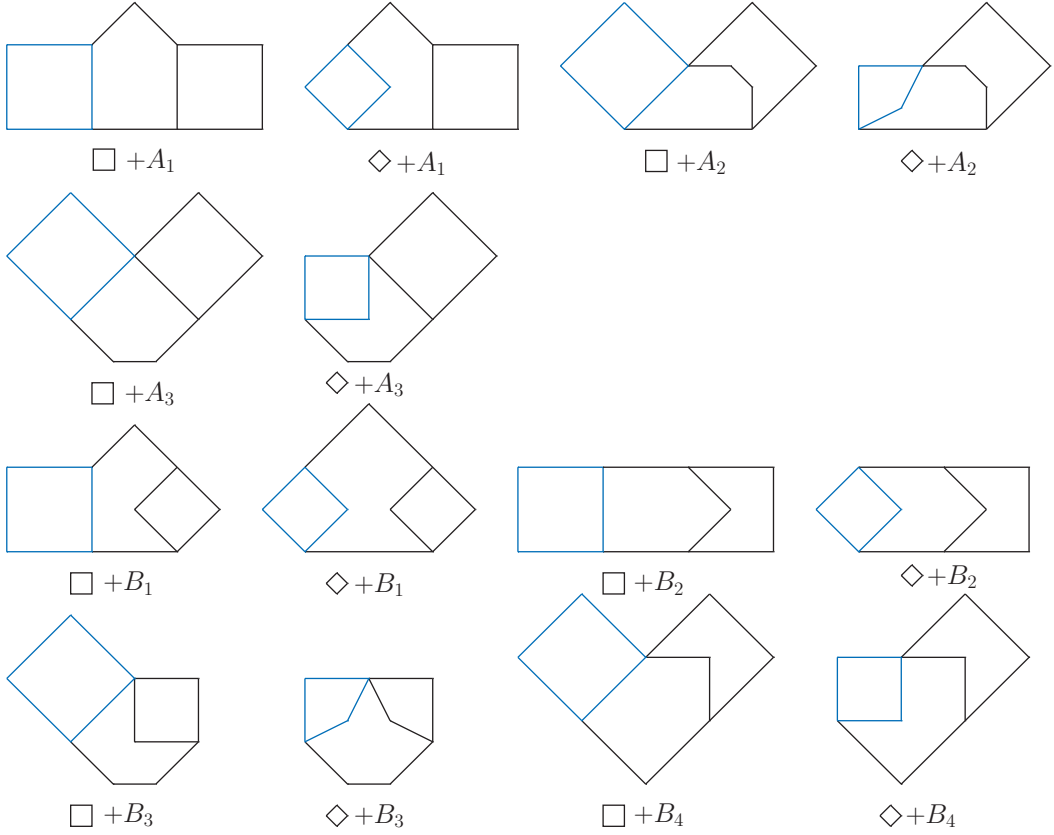}\\
  \caption{Three-loop ladder type diagrams constructed from one-loop box diagram and two-loop double-box, crossed-box
  diagrams. Every vertex is attached by massive external legs, which
  are not explicitly shown in the figure. Diagrams $(\Diamond+A_1)$ and $(\square+B_1)$ are the same,
while diagrams $(\Diamond+A_3)$ and $(\square+B_3)$ are also the
same. So there are in total twelve distinct diagrams.}\label{3LoopT1}
\end{figure}

To finish this subsection, let us present the results for ladder
type diagrams. There are 13 diagrams whose cut equations define
non-trivial curves. Twelve of them have a sub-two-loop double-box or
crossed-box diagram, denoted by $(n_1,n_2,n_3,n_4,n_5,n_6)$ as
\bea
&&(2,2,2,2,2,1)~,~(3,1,1,3,2,1)~,~(3,1,2,2,2,1)~,~(3,2,1,3,1,1)~,~(3,2,2,2,1,1)~,~~~\nonumber\\
&&(2,2,2,2,0,3)~,~(3,1,1,3,0,3)~,~(3,1,2,2,0,3)~,~(3,2,1,3,0,2)~,~(3,2,2,2,0,2)~,~~~\nonumber\\
&&(3,3,1,3,0,1)~,~(3,3,2,2,0,1)~.~~~\nonumber\eea
Genus of these twelve diagrams can be computed by the recursive
formula (\ref{3to2recur1}) or (\ref{3to2recur2}). The construction
of these diagrams are shown in Figure (\ref{3LoopT1}).
With the known results
$g_{\includegraphics[width=1.3em]{dboxSym.eps}}=1$ and
$g_{\includegraphics[width=1em]{cboxSym.eps}}=3$, using formula
(\ref{3to2recur1}), we can compute the genus as
\begin{center}
\begin{tabular}{|c|c|c|c|c|c|c|c|}
  \hline
  % after \\: \hline or \cline{col1-col2} \cline{col3-col4} ...
  ~ & $~~A_1~~$ & $~~A_2~~$ & $~~A_3~~$ & $~~B_1~~$ & $~~B_2~~$ & $~~B_3~~$ & $~~B_4~~$ \\
  \hline
  $\square$ & 5 & 9 & 5 & 9 & 13 & 9 & 13 \\
  \hline
  $\Diamond$ & 9 & 17 & 9 & 13 & 21 & 13 & 21 \\
  \hline
\end{tabular}
\end{center}
Note that diagram $(\Diamond+A_1)$ and $(\square+B_1)$ are the same
diagram, while diagram $(\Diamond+A_3)$ and $(\square+B_3)$ are also
the same diagram.

%%%%%%%%%%%%%%%%%
\subsection{Mercedes-logo type diagrams}
%%%%%%%%%%%%%%%%%
\begin{figure}
  % Requires \usepackage{graphicx}
  \centering
  \includegraphics[width=6in]{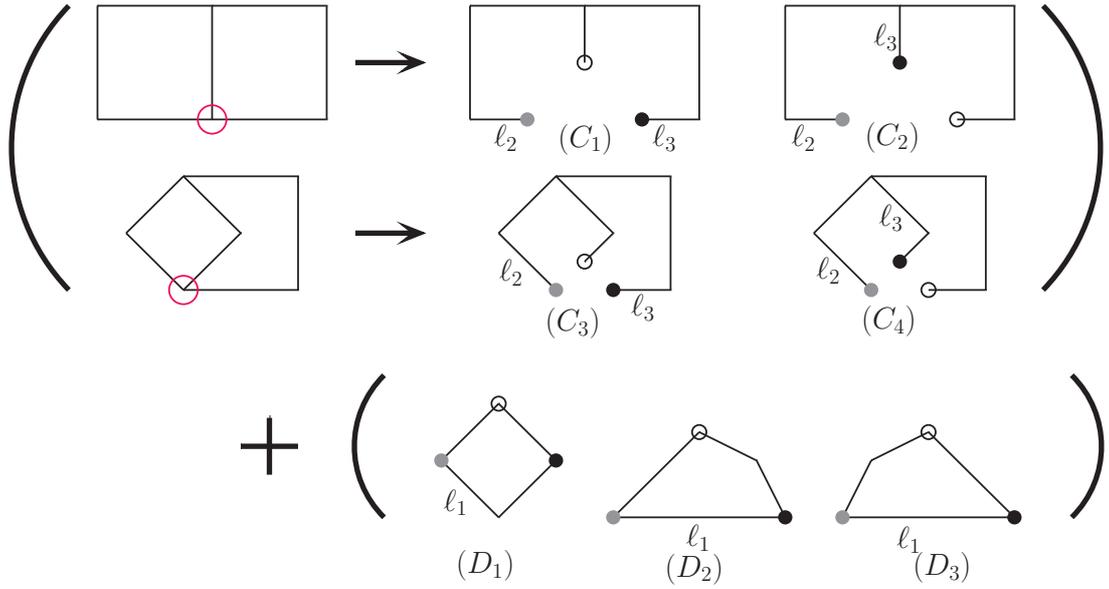}\\
  \caption{Different ways of connecting one-loop box diagram and two-loop double-box, crossed-box
  diagrams to construct three-loop Mercedes-logo type diagrams. Vertices of two-loop diagrams marked as red circles are opened,
   and the internal lines are connected to the one-loop box diagram at the vertices marked as dots, corresponding to the color of dots.}\label{construct3LT2}
\end{figure}
The Mercedes-logo type diagrams can be constructed by inserting
box-diagram into double-box diagram or crossed-box diagram at the
vertices marked as red circles in Figure (\ref{construct3LT2}).
There are four different ways of connecting to the two-loop diagrams
and three different ways of connecting to the box diagram, as shown
in Figure (\ref{construct3LT2}). They are connected at the vertices
marked as dots, corresponding to the color of dots. Discussion on
the equations of sub-two-loop diagram has no difference from ladder
type diagrams. However, equations in the $\mathcal{P}_{\square}$
part become different. There are three quadratic equations
$Q_1(\mathbf{x}^2)=0$, $Q_2(\mathbf{x}\mathbf{y})=0$ and
$Q_3(\mathbf{x}\mathbf{z})=0$, but only one linear equation. The
covering map from Mercedes-logo type diagram to double-box or
crossed-box diagram is then given by
\bea \mathcal{C}_{D_i+C_j}:~\left\{\begin{array}{l}
                                Q_1(\mathbf{x}^2)=0\\
                              Q_2(\mathbf{x}\mathbf{y})=0 \\
                              Q_3(\mathbf{x}\mathbf{z})=0 \\
                              L'_1=0\\
                              Q_4(\mathbf{y}^2)=0 \\
                              Q_5(\mathbf{y}\mathbf{z})=0 \\
                              Q_6(\mathbf{z}^2)=0\\
                              L_1=0\\
                              L_2=0\\
                              L_3=0\\
                              L_4=0
                            \end{array}\right.~~\mapsto~~\mathcal{C}_{\includegraphics[width=1.3em]{dboxSym.eps},\includegraphics[width=1em]{cboxSym.eps}}:~
\left\{\begin{array}{l}
                              Q_4(\mathbf{y}^2)=0 \\
                              Q_5(\mathbf{y}\mathbf{z})=0 \\
                              Q_6(\mathbf{z}^2)=0\\
                              L_1=0\\
                              L_2=0\\
                              L_3=0\\
                              L_4=0
                            \end{array}\right.~.~~~\eea
Since equations $Q_2=Q_3=L'_1=0$ are always linear in $\mathbf{x}$,
there is in fact only one quadratic equation in $\mathbf{x}$, and it
defines two covering sheets over
$\mathcal{C}_{\includegraphics[width=1.3em]{dboxSym.eps},\includegraphics[width=1em]{cboxSym.eps}}$.
For any given point in the curve, equations $Q_1=Q_2=Q_3=L'_1=0$
gives two solutions. Only when the discriminant equals to zero,
these two solutions coincide to each other. In this case the point
$P$ becomes ramified point with ramification index $e_P=2$. In our
convention, $u=v=1$, then the number of ramified points is given by
\bea
N&=&8\big(2+m'_y+m'_z+m_{xy}(m'_{yz}-m'_z)+m_{xz}(m'_{yz}-m'_y)\big)~.~~~\label{MercedesR}\eea
As noted before, at most one of $m_{xy}, m_{xz}$ could be one. If
$m_{xy}=m_{xz}=0$, then $N=8(2+m'_y+m'_z)$. If $m_{xy}=1,m_{xz}=0$,
then $N=8(2+m'_y+m'_{yz})$. Similarly, if $m_{xy}=0,m_{xz}=1$, then
$N=8(2+m'_z+m'_{yz})$. So for given number of each propagators, $N$
could be 16, 24 or 32.

For the sub-two-loop diagram, the four linear equations of four
diagrams in Figure (\ref{construct3LT2}) are given by
\bea \left\{\begin{array}{l}
       L^{C_1}_1(\mathbf{y})=0 \\
       L^{C_1}_2(\mathbf{y})=0 \\
       L^{C_1}_3(\mathbf{z})=0 \\
       L^{C_1}_4(\mathbf{z})=0
     \end{array}\right.~~,~~\left\{\begin{array}{l}
       L^{C_2}_1(\mathbf{y})=0 \\
       L^{C_2}_2(\mathbf{y})=0 \\
       L^{C_2}_3(\mathbf{y},\mathbf{z})=0 \\
       L^{C_2}_4(\mathbf{y},\mathbf{z})=0
     \end{array}\right.~~,~~\left\{\begin{array}{l}
       L^{C_3}_1(\mathbf{y})=0 \\
       L^{C_3}_2(\mathbf{y},\mathbf{z})=0 \\
       L^{C_3}_3(\mathbf{z})=0 \\
       L^{C_3}_4(\mathbf{z})=0
     \end{array}\right.~~,~~\left\{\begin{array}{l}
       L^{C_4}_1(\mathbf{y})=0 \\
       L^{C_4}_2(\mathbf{z})=0 \\
       L^{C_4}_3(\mathbf{y},\mathbf{z})=0 \\
       L^{C_4}_4(\mathbf{y},\mathbf{z})=0
     \end{array}\right.~.~~~\nonumber\eea
So we have
\bea
&&C_1:~m'_y=0~~,~~m'_z=0~~,~~m'_{yz}=1~~~,~~~C_2:~m'_y=0~~,~~m'_z=1~~,~~m'_{yz}=0~,~~~\nonumber\\
&&C_3:~m'_y=1~~,~~m'_z=0~~,~~m'_{yz}=1~~~,~~~C_4:~m'_y=1~~,~~m'_z=1~~,~~m'_{yz}=0~.~~~\nonumber\eea
For the box part, we have
\bea
D_1:~m_{xy}=0~~,~~m_{xz}=0~~~,~~~D_2:~m_{xy}=0~~,~~m_{xz}=1~~~,~~~D_3:~m_{xy}=1~~,~~m_{xz}=0~.~~~\nonumber\eea
With above information, we can simply write down the genus of
Mercedes-logo type diagrams by Riemann-Hurwitz formula. The
recursive formula is given by \iffalse % \bea
2g_{C_i+D_j}-2=2(2g_{\includegraphics[width=1.3em]{dboxSym.eps},\includegraphics[width=1em]{cboxSym.eps}}-2)+8\big(2+m'_y+m'_z+m_{xy}(m'_{yz}-m'_z)+m_{xz}(m'_{yz}-m'_y)\big)~.~~~\eea
%
\fi
%
\bea
g_{C_i+D_j}=2g_{\includegraphics[width=1.3em]{dboxSym.eps},\includegraphics[width=1em]{cboxSym.eps}}-1+4\big(2+m'_y+m'_z+m_{xy}(m'_{yz}-m'_z)+m_{xz}(m'_{yz}-m'_y)\big)~.~~~\eea
\begin{figure}
  % Requires \usepackage{graphicx}
  \centering
  \includegraphics[width=5.5in]{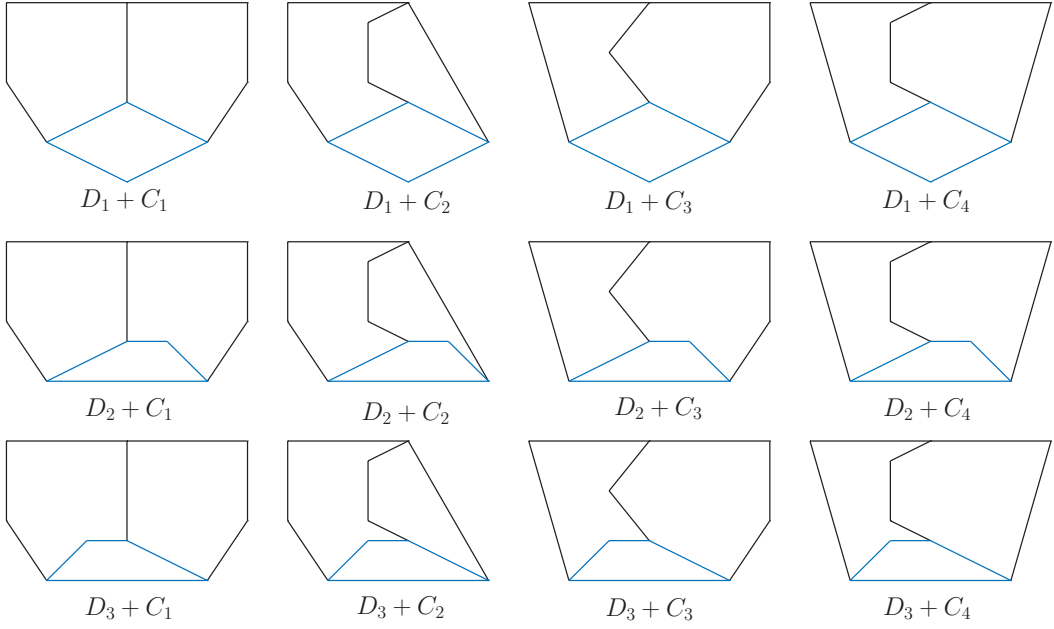}\\
  \caption{Three-loop Mercedes-logo type diagrams constructed from one-loop box diagram and two-loop double-box, crossed-box
  diagrams. Every vertex is attached by massive external legs, which
  are not explicitly shown in the figure. By loop momentum redefinition, there are only four distinct diagrams among the twelve diagrams.}\label{3LoopT2}
\end{figure}

To finish this subsection, we present the results for Mercedes-logo
type diagrams. By naively combining $C_i,D_j$, there are in total
twelve diagrams, as shown in Figure (\ref{3LoopT2}). The genus is
given by
\begin{center}
\begin{tabular}{|c|c|c|c|c|}
  \hline
  % after \\: \hline or \cline{col1-col2} \cline{col3-col4} ...
  ~ & $~~C_1~~$ & $~~C_2~~$ & $~~C_3~~$ & $~~C_4~~$ \\
  \hline
  $D_1$ & 9 & 13 & 17 & 21 \\
  \hline
  $D_2$ & 13 & 13 & 17 & 17 \\
  \hline
  $D_3$ & 13 & 9 & 21 & 17 \\
  \hline
\end{tabular}
\end{center}
However, by loop momenta redefinition, we find that there are in
fact only four different diagrams in Figure (\ref{3LoopT2}), denoted
by $(n_1,n_2,n_3,n_4,n_5,n_6)$ as
\bea
(2,2,3,1,1,2)~,~(2,1,3,2,1,2)~,~(3,2,3,1,1,1)~,~(3,1,3,2,1,1)~.~~~\nonumber\eea
Diagrams with the same genus in above table are the same diagram
after loop momenta redefinition.

%%%%%%%%%%%%%%%%%%%%%%
\subsection{The derivation of formula}
%%%%%%%%%%%%%%%%%%%%

In order to have a general discussion, let us write the eleven
equations of maximal unitarity cuts in a generic form. We always
assume to have already reduced as many equations as possible to
linear equations by algebraic manipulation of performing $D_i-D_j$.

The four equations of box diagram can be expressed as
\bea &&f_1=0=x_1 x_2+y_1 y_2~,~~~\label{f1}\\
&&f_2=0=x_1 y_2^{u}+x_2 y_1^{u}+x_3 y_4^{u}+x_4
y_3^{u}+\sum_{i=1}^{4}a_i(x_i-y_i^{u})+a_0~,~~~\label{f2}\\
&&f_3=0=x_1 z_2^{v}+x_2 z_1^{v}+x_3 z_4^{v}+x_4
z_3^{v}+\sum_{i=1}^{4}b_i(x_i+z_i^{v})+b_0~,~~~\label{f3}\\
&&f_4=0=c_1x_1+c_2x_2+c_3x_3+c_4x_4+w(\mathbf{y}^{m_{xy}},\mathbf{z}^{m_{xz}})+c_0~,~~~\label{f4}\eea
where $u=(n_{xy}-m_{xy})$, $v=(n_{xz}-m_{xz})$ are the number of
quadratic equations containing $\{\mathbf{x},\mathbf{y}\}$ and
$\{\mathbf{x},\mathbf{z}\}$ respectively. Since the box diagram part
contains four propagators, we have $n_x+n_{xy}+n_{xz}=4$. The
function
\bea
w(\mathbf{y}^{m_{xy}},\mathbf{z}^{m_{xz}})=\sum_{i=1}^4c_i(-y_i^{m_{xy}}+z_i^{m_{xz}})~~~~\eea
is a linear function of either $\mathbf{y}$ or $\mathbf{z}$, since
$m_{xy}, m_{xz}$ can take the value of one or zero, but they can not
take the value of one simultaneously. Consequently, equation $f_4=0$
could be a linear function of either $\{\mathbf{x}\}$,
$\{\mathbf{x},\mathbf{y}\}$ or $\{\mathbf{x},\mathbf{z}\}$. Note
that in our convention, there will always be a quadratic equation of
$\{\mathbf{x},\mathbf{y}\}$, so $u\equiv 1$. We keep it undefined
just for generality.

The cut equations for two-loop diagram part can be expressed as
\bea &&g_1=0=y_1 y_2+y_3 y_4~,~~~\label{g1}\\
&&g_2=0=z_1 z_2+z_3z_4~,~~~\label{g2}\\
&&g_3=0=y_1 z_2+y_2z_1+y_3 z_4+y_4
z_3+\sum_{i=1}^{4}d_i(y_i+z_i)+d_0~,~~~\label{g3}\eea
together with other four linear equations $g_4=g_5=g_6=g_7=0$ of
$\{\mathbf{y},\mathbf{z}\}$.

The ramified points are defined by above seven equations of
sub-two-loop part together with the discriminant of $\mathbf{x}$
computed from box diagram part. It is a zero-dimensional ideal, and
always has finite number of point solutions. Let us start from the
analysis of discriminant. By solving three $x_i$'s with equations
$f_2=f_3=f_4=0$, we can write $f_1=0$ as a quadratic equation of
remaining one variable $x_i$. It is simple to compute the
discriminant of this quadratic equation, although the explicit
expression is too tedious to write down. The result takes the
schematic form
\bea
\Delta=h_1(\mathbf{y}^{2u}\mathbf{z}^{2v})+wh_2(\mathbf{y}^{2u}\mathbf{z}^{2v})+w^2h_3(\mathbf{y}^{2u}\mathbf{z}^{2v})~,~~~\label{discri}\eea
where $h_i$'s are generic polynomials of $\{\mathbf{y},\mathbf{z}\}$
with the degree dependence as shown in the argument. Note that we do
not explicitly write down the dependence of lower degree monomials
in $h_i$'s. It is clear that if
\bea
&&v=0~~,~~m_{xy}=m_{xz}=0~~,~~\Delta=\Delta(\mathbf{y}^2)~~\mbox{of~degree}~~2~,~~~\nonumber\\
&&v=0~~,~~m_{xy}~\mbox{or}~m_{xz}=1~~,~~\Delta=\Delta(\mathbf{y}^{2+2m_{xy}}\mathbf{z}^{2m_{xz}})~~\mbox{of~degree}~~4~,~~~\nonumber\\
&&v=1~~,~~m_{xy}=m_{xz}=0~~,~~\Delta=\Delta(\mathbf{y}^{2}\mathbf{z}^{2})~~\mbox{of~degree}~~4~,~~~\nonumber\\
&&v=1~~,~~m_{xy}~\mbox{or}~m_{xz}=1~~,~~\Delta=\Delta(\mathbf{y}^{2+2m_{xy}}\mathbf{z}^{2+2m_{xz}})~~\mbox{of~degree}~~6~.~~~\nonumber\eea
If other equations $g_i=0$ are general, then above information of
$\{\mathbf{y},\mathbf{z}\}$ dependence in $\Delta$ is sufficient to
determine the number of point solutions by convex hull polytope
method. However, given the special form $g_1,g_2,g_3$ in (\ref{g1}),
(\ref{g2}) and (\ref{g3}), there are non-trivial cancelation in
$\Delta$ we need to explore. The cancelation happens when $u=v=1$.
Naively, in this case $h_3(y_{i_1}y_{i_2}z_{j_1}z_{j_2})$ is a
degree four polynomial. All monomials of degree four in $h_3$ are
given by
\bea (y_1z_2+y_2z_1+y_3
z_4+y_4z_3)^2-4(y_1y_2+y_3y_4)(z_1z_2+z_3z_4)~.~~~\nonumber\eea
We can rewrite it as
\bea
g_3^2-2g_3\Big(\sum_{i=1}^{4}d_i(y_i+z_i)+d_0\Big)+\Big(\sum_{i=1}^{4}d_i(y_i+z_i)+d_0\Big)^2-4g_1
g_2~.~~~\nonumber\eea
So if $g_1=g_2=g_3=0$, it reduces to a polynomial of degree two.
Similarly, all monomials of degree three in $h_3$ can be rewritten
as
\bea
\sum_{i=1}^{4}2a_i(z_ig_3-2y_ig_2)+\sum_{i=1}^{4}2b_i(y_ig_3-2z_ig_1)+\mbox{lower~degree~monomial}~.~~~\nonumber\eea
So it can also be reduced to lower degree monomials provided
$g_1=g_2=g_3=0$. In this case,
$h_3(\mathbf{y}^2,\mathbf{yz},\mathbf{z}^2)$ is actually a generic
polynomial of degree two. The same cancelation happens for $h_2$.
All monomials of degree four in function $h_2$ can be rewritten as
\bea
\Big(\sum_{i=1}^{4}a_iy_i\Big)\sum_{i=1}^{4}2c_i(z_ig_3-2y_ig_2)-\Big(\sum_{i=1}^{4}b_iz_i\Big)\sum_{i=1}^{4}2c_i(y_ig_3-2z_ig_1)+\mbox{lower~degree~monomial}~.~~~\nonumber\eea
So when considering $g_1=g_2=g_3=0$, $h_2$ is a generic polynomial
of degree three
$h_2(\mathbf{y}^2\mathbf{z},\mathbf{y}\mathbf{z}^2)$. The
discriminant $\Delta$ can at most be degree four when $w$ is
$\mathbf{y}$ or $\mathbf{z}$-dependent.

A further observation on $f_i$ shows that, the dependence of linear
terms in $f_2, f_3, f_4$ are in fact not arbitrary. For example,
when $u=1$, in the quadratic polynomial $f_2$, the eight linear
terms have only four arbitrary pre-factors $a_i,i=1,2,3,4$, and
$(x_i-y_i)$ always appear together. It is the same for $f_3$ when
$v=1$, $(x_i+z_i)$ will always appear as one single item, and there
are only four arbitrary pre-factors. Also in $f_4$, we always have
$(x_i-y_i^{m_{xy}}+z_i^{m_{xz}})$ as a single item appearing in the
linear equation. This observation leads to non-trivial reformulation
for the discriminant when combined with equations $g_1=g_2=g_3=0$,
while $w$ is $\mathbf{y}$ or $\mathbf{z}$-dependent. More
explicitly, when $u=v=1$ and $m_{xy}=1, m_{xz}=0$, the discriminant
(\ref{discri}) becomes a polynomial of degree four, while the
highest degree of $\mathbf{y}$ is four and the highest degree of
$\mathbf{z}$ is two. If we redefine $z_i=\widetilde{z}_i-y_i$, then
the discriminant can be rewritten as
\bea
\Delta=\sum_{i,j=1}^{4}\widetilde{h}_{1,ij}(\mathbf{y}^2)\widetilde{z}_i\widetilde{z}_j+\sum_{i=1}^{4}\widetilde{h}_{2,i}(\mathbf{y}^3)\widetilde{z}_i+\widetilde{h}_3(\mathbf{y}^4)~.~~~\nonumber\eea
It can be found that
$\widetilde{h}_3(\mathbf{y}^4)=(y_1y_2+y_3y_4)h'_3(\mathbf{y}^2)$,
so it vanishes in case that $g_1=0$. The $\widetilde{h}_{2,i}$ does
not vanish individually, however the summation
$\sum_{i=1}^{4}\widetilde{h}_{2,i}(\mathbf{y}^3)(y_i+z_i)$ vanishes
when combined with the equations $g_1=g_2=g_3=0$. So finally the
discriminant can be expressed as
\bea
\Delta=\sum_{i,j=1}^{4}\widetilde{h}_{1,ij}(\mathbf{y}^2)(y_i+z_i)(y_j+z_j)=\Delta(\mathbf{y}^2(\mathbf{y}+\mathbf{z})^2)~.~~~\nonumber\eea
Similarly, when $m_{xy}=0, m_{xz}=1$, the discriminant can be
expressed as
\bea
\Delta=\sum_{i,j=1}^{4}\widetilde{h}_{1,ij}(\mathbf{z}^2)(y_i+z_i)(y_j+z_j)=\Delta(\mathbf{z}^2(\mathbf{y}+\mathbf{z})^2)~.~~~\nonumber\eea

\begin{figure}
  % Requires \usepackage{graphicx}
  \centering
  \includegraphics[width=5.5in]{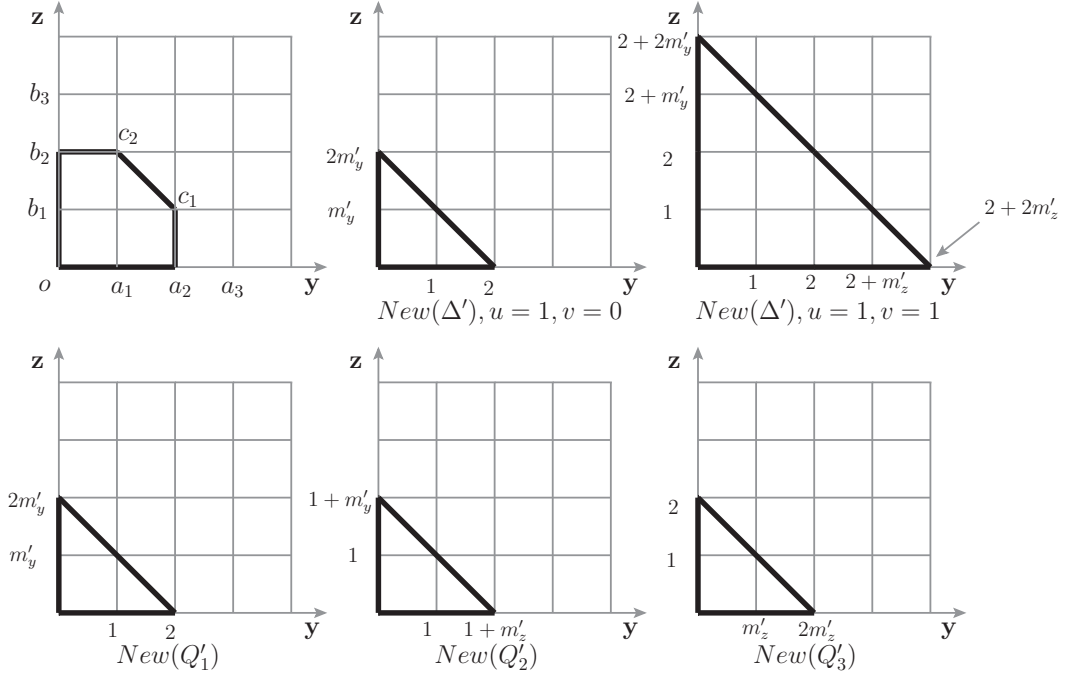}\\
  \caption{Lattice convex polytopes associated with polynomial equations. The coordinate is scaled by $k^2/2$,
  where $k$ is the coordinate of lattice segments. }\label{polytope4D}
\end{figure}

We have explored all the hidden structures in the discriminant
$\Delta$ under given equations $g_1=g_2=g_3=0$ in (\ref{g1}),
(\ref{g2}) and (\ref{g3}). The degree dependence in $\Delta$ is
determined by $u,v$ and $m_{xy},m_{xz}$, and can be summarized as
\bea
\Delta\Big(\mathbf{y}^{2u(1-m_{xz}+(m_{xy}+m_{xz})(1-uv))}(\mathbf{y}+\mathbf{z})^{2uv(m_{xy}+m_{xz})}
\mathbf{z}^{2v((1-m_{xy})+(m_{xy}+m_{xz})(1-uv))} \Big)~.~~~\eea
For any given $u,v,m_{xy},m_{xz}$ from cut equations of box diagram
part, it is a degree four polynomial, and the degree dependence of
$\mathbf{y}, \mathbf{z}$ and $(\mathbf{y}+\mathbf{z})$ is explicitly
shown. We want to emphasize that, $\Delta$ is expressed as the above
form such that at most two terms in $\mathbf{y},\mathbf{z}$ or
$(\mathbf{y}+\mathbf{z})$ could appear at the same time. For all
possible values of $u,v, m_{xy},m_{xz}$ from cut equations, the
discriminant can be
\bea
\Delta(\mathbf{y}^2)~~,~~\Delta(\mathbf{y}^4)~~,~~\Delta(\mathbf{z}^2)~~,~~\Delta(\mathbf{z}^4)~~,~~
\Delta(\mathbf{y}^2(\mathbf{y}+\mathbf{z})^2)~~,~~\Delta(\mathbf{z}^2(\mathbf{y}+\mathbf{z})^2)~~.~~\nonumber\eea
Then it is possible to compute the mixed volume of polytopes defined
by polynomials $\Delta, g_i,i=1,\ldots,7$. Naively, these
polynomials are associated with 8-dimensional polytope, and it is
not easy to compute the 8-dimensional volume. However, since there
are four linear equations, we can solve four variables, and the
remaining four equations are associated with 4-dimensional volume.
It is still not easy to compute arbitrary 4-dimensional volume. But
if we always choose to solve $\{y_3,y_4,z_3,z_4\}$ from four liner
equations, then the remaining four variables
$\{\mathbf{y}_s,\mathbf{z}_s\}=\{y_1,y_2,z_1,z_2\}$ are symmetric
among $\{y_1,y_2\}$ or $\{z_1,z_2\}$. Then we can treat
$\mathbf{y}_s$ or $\mathbf{z}_s$ as a lattice line whose segment
coordinate equals to the area of triangle in $\{y_1,y_2\}$ or
$\{z_1,z_2\}$-plane. An example is shown in the first diagram of
Figure~(\ref{polytope4D}).
The length $\overline{oa_1}=\overline{ob_1}={1^2\over 2}$,
$\overline{oa_2}=\overline{ob_2}={2^2\over 2}$,
$\overline{oa_3}=\overline{ob_3}={3^2\over 2}$, etc. Instead of
computing the 4-dimensional volume directly, we compute the
2-dimensional area but with a scaled coordinate. The polytope shown
in the first diagram of Figure (\ref{polytope4D}) then has area
\bea &&{\overline{oa_2}+\overline{b_1c_1}\over 2}\times
\overline{ob_1}+{\overline{b_1c_1}+\overline{b_2c_2}\over 2}\times
(\overline{ob_2}-\overline{ob_1})\nonumber\\
&&={2^2/2+2^2/2\over 2}\times {1^2\over 2}+{2^2/2+1^2/2\over
2}\times({2^2\over 2}-{1^2\over 2})={23\over 8}~.~~~\nonumber\eea

Special attention should be paid to the case when discriminant is
given by $\Delta(\mathbf{y}^2(\mathbf{y}+\mathbf{z})^2)$ or
$\Delta(\mathbf{z}^2(\mathbf{y}+\mathbf{z})^2)$. Because of the
dependence of $(\mathbf{y}+\mathbf{z})$, we should treat
$(\mathbf{y}+\mathbf{z})$ as a variable. So when
$\Delta=\Delta(\mathbf{y}^2(\mathbf{y}+\mathbf{z})^2)$, we should
transform the variables $\mathbf{z}\to
\widetilde{\mathbf{z}}-\mathbf{y}$ such that the discriminant become
$\Delta(\mathbf{y}^2\widetilde{\mathbf{z}}^2)$. Similarly, when
$\Delta=\Delta(\mathbf{z}^2(\mathbf{y}+\mathbf{z})^2)$, we should
transform the variables $\mathbf{y}\to
\widetilde{\mathbf{y}}-\mathbf{z}$ such that the discriminant become
$\Delta(\widetilde{\mathbf{y}}^2\mathbf{z}^2)$. Then we can compute
the 4-dimensional mixed volume accordingly. Let us take
$m_{xy}=m_{xz}=0$ for example. In this case the discriminant is
$\Delta(\mathbf{y}^{2u(1-m_{xz})}\mathbf{z}^{2v(1-m_{xy})})$, so we
do not need to transform variables. The solution of linear equations
can be formally written as
\bea \mathbf{y}\mapsto
\mathbf{y}_s~,~\mathbf{z}_s^{m'_y}~~~,~~~\mathbf{z}\mapsto
\mathbf{z}_s~,~\mathbf{y}_s^{m'_z}~.~~~\nonumber\eea
In this case, the three quadratic equation $Q(\mathbf{y}^2)$,
$Q(\mathbf{yz})$ and $Q(\mathbf{z}^2)$ become
\bea Q'_1(\mathbf{y}_s^2,\mathbf{y}_s
\mathbf{z}_s^{m'_y},\mathbf{z}_s^{2m'_y})~~,~~Q'_2(\mathbf{y}_s^{1+m'_z},\mathbf{y}_s\mathbf{z}_s,\mathbf{z}_s^{1+m'_y})
~~,~~Q'_3(\mathbf{z}_s^2,
\mathbf{y}_s^{m'_z}\mathbf{z}_s,\mathbf{y}_s^{2m'_z})~.~~~\nonumber\eea
The discriminant can be expressed as
\bea
\Delta'(\mathbf{y}_s^{2u+2vm'_z},\mathbf{z}_s^{2v+2um'_y})~.~~~\nonumber
\eea
The polytopes associated with these polynomials are plotted in
Figure (\ref{polytope4D}). $New(\Delta')$ is drawn explicitly with
given $u,v$ for computation purpose, and the coordinate of vertices
of polytopes are marked along the axes. Although these polytopes are
plotted universally as triangles, we should note that they depend on
the value of $m'_y, m'_z$. For example, if $m'_y=0$, $New(Q'_2)$ is
a trapezoid. Given the four polytopes $New(\Delta')$, $New(Q'_1)$,
$New(Q'_2)$ and $New(Q'_3)$ with their coordinates, it is
straightforward to draw the Minkowski sum among them. Then we can
compute the mixed volume according to formula (\ref{mixedvolume4d}).
We find that the mixed volume for $m_{xy}=m_{xz}=0$ is given by
\bea \mathcal{M}(\Delta', Q'_1, Q'_2,
Q'_3)=8(u+v+um'_y+vm'_z)~.~~~\eea
Similarly, when $m_{xy}=1, m_{xz}=0, u=v=1$, we have variables
$\mathbf{y},\widetilde{\mathbf{z}}$. The solution of linear
equations are given by
\bea \mathbf{y}\mapsto
\mathbf{y}_s~,~\mathbf{\widetilde{z}}_s^{m'_{y}}~~~,~~~\mathbf{\widetilde{z}}\mapsto
\mathbf{\widetilde{z}}_s~,~\mathbf{y}_s^{m'_{yz}}~.~~~\nonumber\eea
In this case, we have
\bea Q'_1(\mathbf{y}_s^2,\mathbf{y}_s
\mathbf{\widetilde{z}}_s^{m'_y},\mathbf{\widetilde{z}}_s^{2m'_y})~~,~~Q'_2(\mathbf{y}_s^{1+m'_{yz}},\mathbf{y}_s\mathbf{\widetilde{z}}_s,\mathbf{\widetilde{z}}_s^{1+m'_y})
~~,~~Q'_3(\mathbf{\widetilde{z}}_s^2,
\mathbf{y}_s^{m'_{yz}}\mathbf{\widetilde{z}}_s,\mathbf{y}_s^{2m'_{yz}})~,~~~\nonumber\eea
and
$\Delta'(\mathbf{y}_s^{2+2m'_{yz}},{\mathbf{\widetilde{z}}_s}^{2+2m'_y})$.
So the same computation shows that the mixed volume of four
polytopes is given by
\bea \mathcal{M}(\Delta', Q'_1, Q'_2,
Q'_3)=8(2+m'_y+m'_{yz})~.~~~\eea
Finally, if $m_{xz}=1, m_{xy}=0, u=v=1$, we have variables
$\mathbf{z},\widetilde{\mathbf{y}}$. The solution of linear
equations is given by
\bea \mathbf{\widetilde{y}}\mapsto
\mathbf{\widetilde{y}}_s~,~\mathbf{z}_s^{m'_{yz}}~~~,~~\mathbf{z}\mapsto
\mathbf{z}_s~,~\mathbf{\widetilde{y}}_s^{m'_{z}}~~.~~~\nonumber\eea
In this case, we have
\bea Q'_1(\mathbf{\widetilde{y}}_s^2,
\mathbf{z}_s^{m'_{yz}}\mathbf{\widetilde{y}}_s,\mathbf{z}_s^{2m'_{yz}})~~,~~Q'_2(\mathbf{z}_s^{1+m'_{yz}},\mathbf{z}_s\mathbf{\widetilde{y}}_s,\mathbf{\widetilde{y}}_s^{1+m'_z})
~~,~~Q'_3(\mathbf{z}_s^2,\mathbf{z}_s
\mathbf{\widetilde{y}}_s^{m'_z},\mathbf{\widetilde{y}}_s^{2m'_z})~,~~~\nonumber\eea
and
$\Delta'(\mathbf{\widetilde{y}}_s^{2+2m'_z},\mathbf{z}_s^{2+2m'_{yz}})$.
Then we get
\bea \mathcal{M}(\Delta', Q'_1, Q'_2,
Q'_3)=8(2+m'_z+m'_{yz})~.~~~\eea
Summarizing above discussions, we can express the number of ramified
points, which equals to the mixed volume of four polytopes, as
\bea N&=&{8u\big(1-m_{xz}+(m_{xy}+m_{xz})(1-uv)\big)(1+m'_y)}\nonumber\\
&&+{8uv(m_{xy}+m_{xz})(1+m'_{yz})}+{8v\big(1-m_{xy}+(m_{xy}+m_{xz})(1-uv)\big)(1+m'_z)}~,~~~\nonumber\eea
which has already been shown in the beginning of this section.

%%%%%%%%%%%%%%%%%5
\section{More diagrams}
%%%%%%%%%%%%%%%%%%%%%
%%%%%%%%%%%%%%%
\subsection{The other three-loop diagrams}
%%%%%%%%%%%%%%%

%
\begin{figure}
  % Requires \usepackage{graphicx}
\centering
  \includegraphics[width=5in]{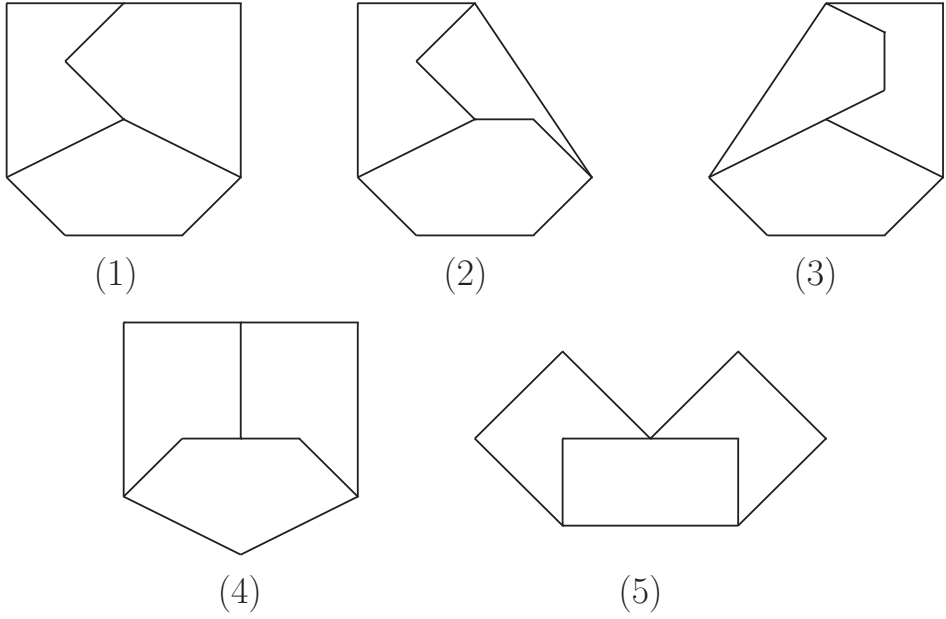}\\
  \caption{Remaining three-loop diagrams whose equations of maximal unitarity cuts define
  curves while their sub-two-loop diagrams do not define curves. Every vertex is attached by massive external legs, which
  are not explicitly shown in the figure.}\label{3LoopEx}
\end{figure}
In previous section, we have presented a recursive formula for the
study of genus of three-loop diagrams whose sub-two-loop diagram
also defines a curve. There are still five diagrams which can not be
included in this category. They are four Mercedes-logo type diagrams
as shown in Figure (\ref{3LoopEx}.1) to Figure (\ref{3LoopEx}.4),
and one ladder type diagram as shown in Figure (\ref{3LoopEx}.5).
Since the sub-two-loop diagram or sub-one-loop diagram does not
define curve, there is no covering map from the original curve to
the curve of lower-loop diagram. Because of the highly complexity of
algebraic system, it is quite difficult to compute the genus
directly. Thus we introduce an algorithm to systematically study the
genus based on numerical algebraic geometry. Given an algebraic
system of maximal unitarity cuts of three-loop diagrams with
arbitrary setup of numeric external momenta, it is possible to
compute the genus within seconds by this algorithm. It also provides
an opportunity of studying the global structure of maximal unitarity
cuts of four-loop and even higher loop diagrams, where analytic
study is almost impossible.

Let us apply the algorithm to the computation of five three-loop
diagrams considered in this subsection. For each diagram, the
corresponding polynomial system of maximal unitarity cuts defines an
irreducible curve $\mathcal{C}_i$ with each contribution
$\rho_{\pi(b)} = 1$ for every $b\in B_{\mathcal{C}_i}$ and
$\rho_{\infty} = 0$. Thus, Riemann-Hurwitz formula reduces to
$$g_{\mathcal{C}_i} =- \deg [\mathcal{C}_i]+1 +
{|B_{\mathcal{C}_i}|\over 2} ~.$$ By computing the degree of
the curve and the number of branchpoints using numerical algebraic geometry
via {\tt Bertini} \cite{Bertini}, we can
obtain the genus by above formula,
\begin{itemize}

\item For diagram (\ref{3LoopEx}.1), we have
$\mbox{deg}[\mathcal{C}]=44$, and $|B_{\mathcal{C}}|=152$, so
the genus is $g=33$.

\item For diagram (\ref{3LoopEx}.2), we also have
$\mbox{deg}[\mathcal{C}]=44$, but $|B_{\mathcal{C}}|=176$, so
the genus is $g=45$.

\item For diagram (\ref{3LoopEx}.3), we have
$\mbox{deg}[\mathcal{C}]=40$, and $|B_{\mathcal{C}}|=136$, so
the genus is $g=29$.

\item Diagram (\ref{3LoopEx}.4) has the most complicated global
structure among three-loop diagrams. The curve associated with
this diagram has degree $\mbox{deg}[\mathcal{C}]=52$, and
$|B_{\mathcal{C}}|=212$, so the genus is $g=55$.

\item For the last ladder type diagram (\ref{3LoopEx}.5), we have
$\mbox{deg}[\mathcal{C}]=32$, and $|B_{\mathcal{C}}|=128$, so
the genus is $g=33$.
\end{itemize}

%%%%%%%%%%%%%%%%%%%%%
\subsection{The White-house diagram}
%%%%%%%%%%%%%%%%%%%%%%
\begin{figure}
  % Requires \usepackage{graphicx}
  \includegraphics[width=6in]{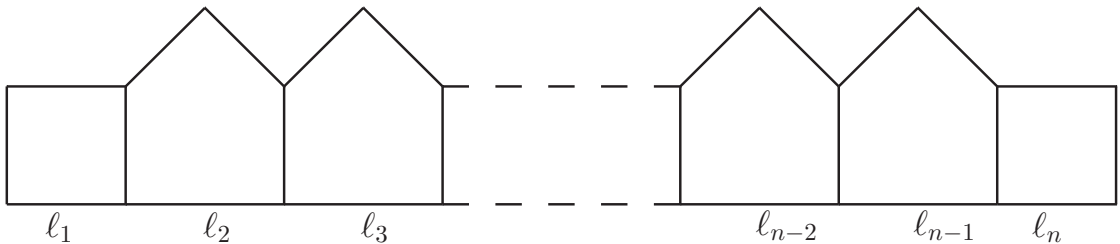}\\
  \caption{Infinite series of white-house diagrams. Every vertex is attached by massive external legs, which
  are not explicitly shown in the figure.}\label{whitehouse}
\end{figure}
An interesting series of diagrams is shown in Figure
(\ref{whitehouse}) to any loop orders. If $n=1$, we get the one-loop
triangle diagram, which has $g^{WH}_{1}=0$. If $n=2$, we get the
two-loop double-box diagram, which has $g^{WH}_{2}=1$. The
three-loop diagram is the first diagram resembling the White-house,
and it has $g^{WH}_3=5$. Because of its resemblance between
$(n-1)$-loop and $n$-loop diagrams, it is interesting to ask if we
can compute the genus of $n$-loop diagram from the information of
$(n-1)$-loop diagram. Define
$\mathbf{x_k}=\{x_{k1},x_{k2},x_{k3},x_{k_4}\}$ as the
parametrization variables of the $k$-th loop. From Figure
(\ref{whitehouse}) we can see that there are always two linear
equations for each loop, so we can solve two variables in
$\mathbf{x_k}$ using these linear equations and get
$\mathbf{x_k}\mapsto \mathbf{x'_k}$, where
$\mathbf{x'_k}=\{x_{k1},x_{k2}\}$. Then we can compute the genus of
$n$-loop white-house diagram by Riemann-Hurwitz formula from the
covering~map
\bea \mathcal{C}^{WH}_{n}:~\left\{\begin{array}{l}
                             Q_1(\mathbf{x_1}^2)=0 \\
                             Q_2(\mathbf{x_1x'_2})=0 \\
                             L_1(\mathbf{x_1})=0 \\
                             L_2(\mathbf{x_1})=0 \\
                             + \\
                             \mathcal{C}^{WH}_{n-1}
                           \end{array}\right.~~\mapsto~~\mathcal{C}^{WH}_{n-1}:~\left\{\begin{array}{l}
                                                                                  Q_3(\mathbf{x'_2}^2)=0 \\
                                                                                  Q_4(\mathbf{x'_2x'_3})=0 \\
                                                                                  ... \\
                                                                                  Q_{2n-3}(\mathbf{x'_{n-1}}^2)=0 \\
                                                                                  Q_{2n-2}(\mathbf{x'_{n-1}}\mathbf{x'_{n}})=0 \\
                                                                                  Q_{2n-1}(\mathbf{x'_n}^2)=0
                                                                                \end{array}\right.~.~~~\eea
As usual, equations $Q_1=Q_2=L_1=L_2=0$ of box diagram part define a
double covering map, and the ramified points have ramification index
$e_P=2$, determined by the discriminant equation $\Delta=0$ for
given points in curve $\mathcal{C}^{WH}_{n-1}$. For the box diagram
part, since $m_{xy}=m_{xz}=0$, $u=1,v=0$, the discriminant is given
by $\Delta(\mathbf{x'_2}^2)$. So the ramified points are determined
by $(2n-2)$ equations in $(2n-2)$ variables in $\mathbb{C}^{2n-2}$.
It is easy to compute the number of ramified points for white-house
diagrams. Since $\Delta(\mathbf{x'_2}^2)=Q_3(\mathbf{x'_2}^2)=0$ are
two generic quadratic equations in two variables
$\mathbf{x'_2}=\{x_{21},x_{22}\}$, by B\'ezout's theorem, it has
four distinct solutions. For each solution
$\{x^{S_i}_{21},x^{S_i}_{22}\}$, equations
$Q_4(\mathbf{x'_2x'_3})=Q_5(\mathbf{x'_3}^2)=0$ are two generic
equations in $\mathbf{x'_3}$ of degree one and two, so they have two
solutions in $\{x_{31},x_{32}\}$. In total we get $4\times 2=8$
solutions in $\{x_{21},x_{22},x_{31},x_{32}\}$. Recursively, we get
$4\times 2^{n-2}=2^n$ distinct solutions in $\{x_{21},x_{22},\ldots,
x_{n1},x_{n2}\}$. Above argument is based on the facts that
different loops only share common propagators adjacently in a chain
and the solution of linear equations only maps $\mathbf{x_k}\mapsto
\mathbf{x'_k}$ itself. So the Riemann-Hurwitz is given by
\bea 2g^{WH}_{n}-2=2(2g^{WH}_{n-1}-2)+2^n~.~~~\eea
Given the first entry $g^{WH}_1=0$, it is not hard to solve above
recursive formula and get
\bea g^{WH}_n=(n-2)2^{n-1}+1~.~~~\eea
It indeed produces $g^{WH}_1=0$, $g^{WH}_2=1$, $g^{WH}_3=5$, and
also an infinite series of genus such as $g^{WH}_4=17$,
$g^{WH}_5=49$, $g^{WH}_6=129$, etc. The genus grows exponentially to
infinity with the increasing of loops, which indicates the
complexity of computation in higher-loop amplitudes.

%%%%%%%%%%%%%%%%%%%
\section{Conclusion}
%%%%%%%%%%%%%%%%%%%%

To systematically study integrand reduction or integral
reduction of multi-loop amplitudes via algebraic geometry method,
the equations derived from propagators on-shell
and their correspoinding variety (solution space) plays a very important role. These on-shell equations are
the generating equations of ideal and Gr\"{o}bner basis, which are
the central objects in determining the set of independent integrand
basis. The solution of each irreducible component of reducible ideal
determines the parametrization of loop momenta, which greatly
affects the evaluation of coefficients for integrand or integral
basis. Thus, the study on the global structure of on-shell equations
is the first step in the process of multi-loop reduction method, and
provides a birds-eye view for further explicit computation.

In order to explicitly apply algebraic geometry method to the
computation of three-loop integrand or integral reduction, it is
necessary as an initial step to elaborate the global structure of
the on-shell equations. Since a four-dimensional three-loop integral
has twelve parametrization variables for loop momenta, the first
category of non-trivial varieties is defined by on-shell equations
of three-loop diagrams with eleven propagators. The ideal defined by
these diagrams is complex one-dimensional, and it defines an
algebraic curve, which is topological equivalent to a Riemann
surface. The global structure is completely characterized by the
geometric genus of the curve. Since these diagrams are the simplest
three-loop diagrams with non-trivial solution space of on-shell
equations, they would be the first candidate for applying algebraic
geometric methods to the explicit computation of three-loop diagrams.
Thus, a thorough study on the global structure of three-loop diagrams
with eleven propagators can be served as a seed for the further
computation of three-loop integrand and integral reduction.

In this paper, we provide a systematic study on the genus of curves
defined by maximal unitarity cuts of three-loop diagrams with eleven
propagators, generalizing the research in  \cite{Huang:2013kh}.
The Riemann-Hurwitz formula is used throughout the study.
Among the 21 total diagrams. 16 diagrams have a sub-two-loop
diagram whose equations of maximal unitarity cut also define curves.
For these diagrams, the genus can be recursively computed from the
genus of two-loop double-box and crossed-box diagrams together with
the knowledge of ramified points. The recursive formula is given by
\bea g=2g_{\includegraphics[width=1.3em]{dboxSym.eps},
\includegraphics[width=1em]{cboxSym.eps}}-1+{N\over 2}~,~~~\eea
where
\bea N&=&{8u\big(1-m_{xz}+(m_{xy}+m_{xz})(1-uv)\big)(1+m'_y)}\nonumber\\
&&+{8uv(m_{xy}+m_{xz})(1+m'_{yz})}+{8v\big(1-m_{xy}+(m_{xy}+m_{xz})(1-uv)\big)(1+m'_z)}~.~~~\nonumber\eea
Note that above formula is general and independent of the convention
of loop momenta. The number $N$ can be obtained by counting the number of
corresponding propagators. Thus, the genus can be evaluated by
just looking at the diagrams.

Besides, there are still five diagrams which can not be analyzed by
the recursive formula from information of two-loop diagrams. For
these diagrams, we implement an algorithm for the Riemann-Hurwitz
formula based on numerical algebraic geometry. This algorithm also
provides the possibility of studying more complicated algebraic
system of four-loop or even higher loop diagrams in the future. It can
also be applied to the analysis of the previous 16 diagrams, and we
find that the results of the recursive formula and the
algorithm~agree.

The genus of 13 ladder type diagrams is given by
\begin{center}
\begin{tabular}{|c|c|c|c|c|c|c|c|c|c|c|c|c|c|}
  \hline
  % after \\: \hline or \cline{col1-col2} \cline{col3-col4} ...
  Diagram & $\includegraphics[width=1.5em]{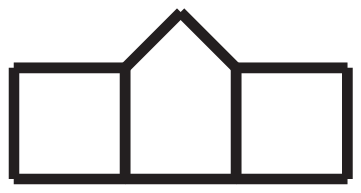}$ & $\includegraphics[width=1.5em]{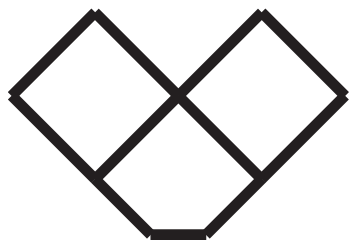}$ & $\includegraphics[width=1.5em]{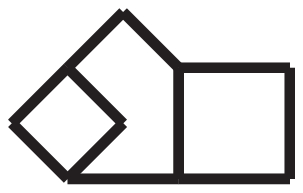}$ & $\includegraphics[width=1.5em]{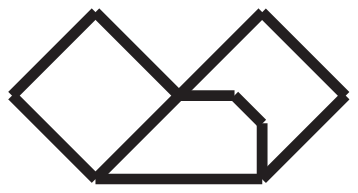}$ & $\includegraphics[width=1.5em]{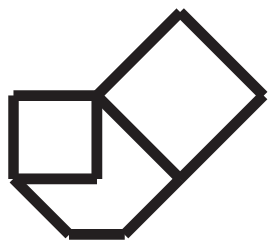}$ & $\includegraphics[width=1.5em]{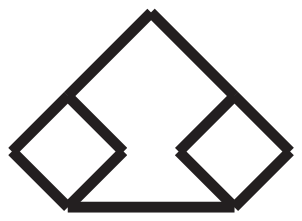}$ & $\includegraphics[width=1.5em]{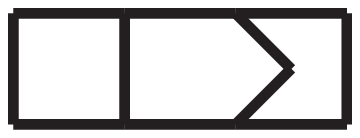}$ & $\includegraphics[width=1.5em]{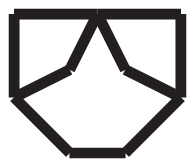}$ & $\includegraphics[width=1.5em]{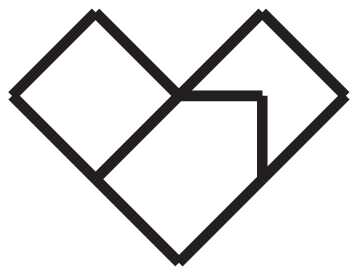}$ & $\includegraphics[width=1.5em]{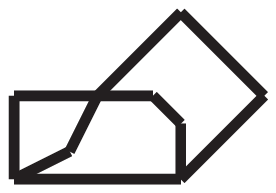}$ & $\includegraphics[width=1.5em]{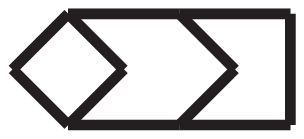}$ & $\includegraphics[width=1.5em]{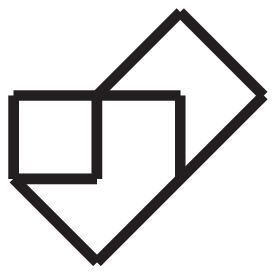}$ & $\includegraphics[width=1.5em]{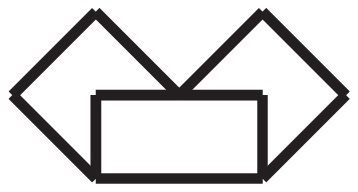}$ \\
  \hline
  Genus & 5 & 5 & 9 & 9 & 9 & 13 & 13 & 13 & 13 & 17 & 21 & 21 & 33 \\
  \hline
\end{tabular}
\end{center}
The genus of 8 Mercedes-logo type diagrams is given by
\begin{center}
\begin{tabular}{|c|c|c|c|c|c|c|c|c|}
  \hline
  % after \\: \hline or \cline{col1-col2} \cline{col3-col4} ...
  Diagram & $\includegraphics[width=1.5em]{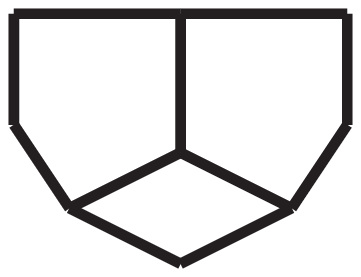}$ & $\includegraphics[width=1.5em]{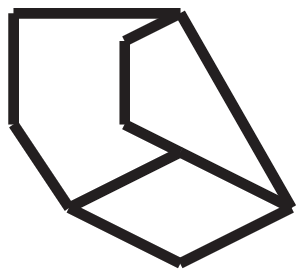}$ & $\includegraphics[width=1.5em]{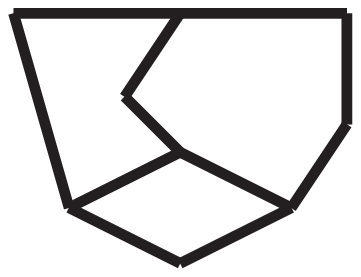}$ & $\includegraphics[width=1.5em]{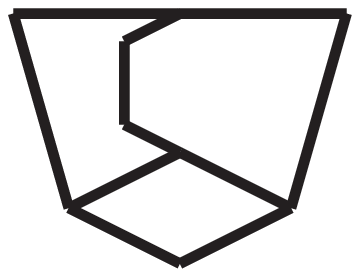}$ & $\includegraphics[width=1.5em]{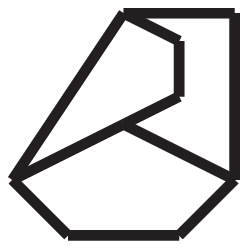}$ & $\includegraphics[width=1.5em]{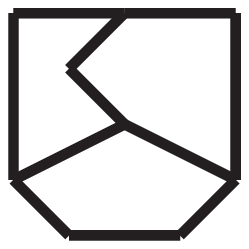}$ & $\includegraphics[width=1.5em]{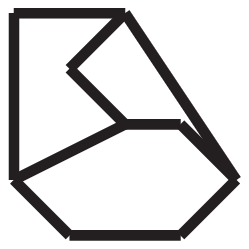}$ & $\includegraphics[width=1.5em]{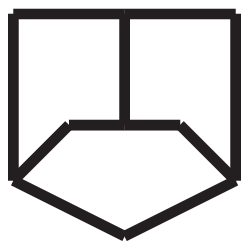}$ \\
  \hline
  Genus & 9 & 13 & 17 & 21 & 29 & 33 & 45 & 55 \\
  \hline
\end{tabular}
\end{center}
In general terms, the higher the genus is,
the more complicated the algebraic system
will be. So, a first direct application of above result would be the
judgement of the complexity of three-loop diagrams we would like to
evaluate.  Different from two-loop diagrams where the highest genus
is only three, the genus for three-loop diagrams can be
as high as 55. This indicates highly complex nature of three-loop diagrams
compared to two-loop diagrams. Curves of different diagrams with the
same genus are topological equivalent to each other, we expect that
this equivalence would also play a role in relating those different
diagrams.

We also present an example beyond three-loop diagrams, by
generalizing the simplest $g=5$ white-house diagram
$\includegraphics[width=1.5em]{A1S.eps}$ to any loops. The genus of
$n$-loop white-house diagram is
\bea g^{WH}_n=(n-2)2^{n-1}+1~.~~~\nonumber\eea
Hence, $g^{WH}_3=5$, $g^{WH}_4=17$, $g^{WH}_5=49$, etc.
In particular, it is possible for the genus to grow without bound.
Interestingly, the genus of $n$-loop white-house diagram equals to the genus of $(n+2)$-dimensional hypercube. This relates the algebraic system of
maximal unitarity cuts of multi-loop diagrams directly to well-known
geometric objects.

An interesting phenomenon is observed in  \cite{Huang:2013kh} that
the genus is always an odd integer. This is further verified by
results presented here. We can claim that the genus of curve defined
by maximal unitarity cuts of any multi-loop diagrams is an odd
integer. This can be shown by looking at the on-shell equations of
degenerate limit where one external momentum is massless. Assuming
that after solving linear equations, the algebraic curve is given by
$I=\langle Q_1,Q_2,\ldots, Q_n\rangle$ with $(n+1)$ variables. It is
always possible to take one external momentum as massless, and the
corresponding quadratic equation factorizes as $Q_1=f_1f_2=0$, where
$f_1,f_2$ are linear. The linear polynomials
$f_1,f_2$ are two equivalent branches of
quadratic polynomial $Q_1$, and ideal $I$ can be primary decomposed
into two equivalent irreducible ideals $I_1=\langle f_1,Q_2,\ldots,
Q_n\rangle$ and $I_2=\langle f_2,Q_2,\ldots,Q_n\rangle$. So, the
genus $g_1$ of
curve defined by $I_1$ equals to the genus of the curve defined by
$I_2$. If there are $N$ intersecting points between two curves, then
the genus of original curve defined by $I$ is given by $g=2g_1+N-1$.
The number $N$ is in fact the number of distinct solutions of
zero-dimensional ideal $I'=\langle f_1,f_2,Q_2,\ldots, Q_n\rangle$.
In projective space, according to B\'ezout's theorem, $N$ is given
by the products of degree of each polynomial in $I'$, which is an
even integer. This guarantees that the genus is an odd
integer, and explains~the~puzzle~in~\cite{Huang:2013kh}.

As we have mentioned, information of global structure of on-shell
equations is the first step to the integrand or integral reduction
of multi-loop integrals. The genus is a powerful concept that
connects the algebraic system of maximal unitarity cuts to geometric
objects, as hinted in the White-house example. Since so far only a
few explicit computation of three-loop integral reduction is done,
we still need to wait for more three-loop examples to reveal the
connection and also the possible equivalence of different diagrams
with the same genus. With the algorithm based on numerical algebraic
geometry presented in this paper, it is possible to work out the
global structure of curves defined by maximal unitarity cuts of
four-loop diagrams. However, this information is not urgent, since
integrand or integral reduction of four-loop integral is still far
from practice. We hope that in future there will be more results of
three-loop integrand reduction showing up, so that we can clarify
the underlining power of genus. Then it can be similarly generalized
to higher loop diagrams.

The global structure of higher-dimensional varieties,
defined by the maximal unitarity cuts of $L$-loop diagram
with $n\leq (4L-2)$ propagators,
is still unclear even for two-loop diagrams. This information is
important for the computation of two-loop diagrams besides
double-box and crossed-box. We hope that both computational
and numerical algebraic geometry can
play a similar role in the analysis of global structure for those diagrams
in the future.

\acknowledgments

We would like to thank Simon Caron-Huot, David Kosower, and Michael
Stillman for useful discussion. RH would like to thank the Niels
Bohr International Academy and Discovery Center, the Niels Bohr
Institute for its hospitality. The work of YZ is supported by Danish
Council for Independent Research (FNU) grant 11-107241. RH's
research is supported by the European Research Council under
Advanced Investigator Grant ERC-AdG-228301. DM was supported by a
DARPA YFA. JDH was supported by a DARPA YFA and NSF DMS-1262428.

\appendix

\section{Solving polynomial equations using convex polytope}

\begin{figure}
  % Requires \usepackage{graphicx}
  \centering
  \includegraphics[width=5.5in]{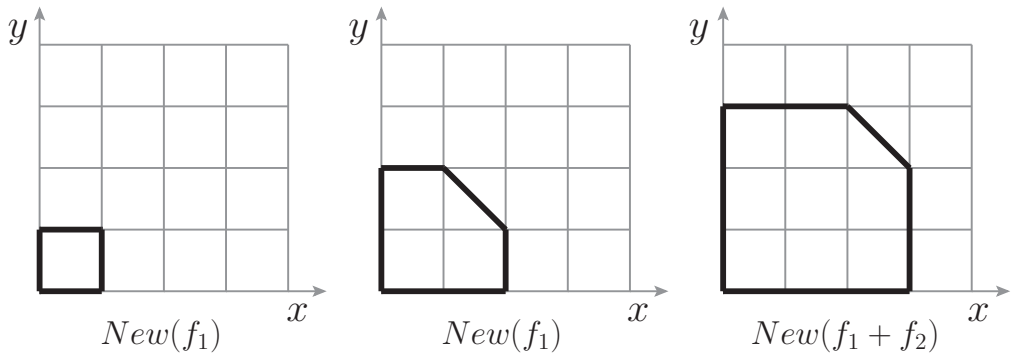}\\
  \caption{Lattice convex polytopes associated with polynomials, and the computation of mixed area.
  Each segment of lattice is length 1.}\label{polyesp}
\end{figure}

An algebraic system of $n$ polynomial equations in $n$ variables
is expected to define a zero-dimensional ideal.
When $n=2$, {\sl if the algebraic system has finite
many zeros in $\mathbb{C}^2$,
then B\'ezout's theorem states that the number of
zeros is at most $\mbox{deg}[f_1]\mbox{deg}[f_2]$}. Generalization
to arbitrary $n$ polynomial equations can be similarly understood.
If there are finitely many zeros in $\mathbb{C}^n$ for
$f_i=0,i=1,\ldots,n$, then the upper bound on the number of
solutions is $\prod_{i=1}^n \mbox{deg}[f_i]$,
which is sharp for generic polynomials.
For sparse polynomials, this bound is typically not sharp.
For illustration, we take a similar example given in
 \cite{mathbookSB}. The~two~polynomials
\bea
f_1(x,y)=a_1+a_2x+a_3xy+a_4y~~,~~~f_2(x,y)=b_1+b_2x^2y+b_3xy^2+b_4x^2+b_5
y^2~~~~~\label{PolyExample}\eea
have four distinct zeros in $\mathbb{C}^2$ for generic coefficients
$a_i,b_i$'s. However, B\'ezout's theorem provides an upper bound of
$\mbox{deg}[f_1]\mbox{deg}[f_2]=6$. In order to predict the actual
number 4 instead of 6, we need to go from B\'ezout's theorem to
Bernstein's theorem. {\sl Bernstein's theorem states that for
two bivariate polynomials $f_1$ and $f_2$,
the number of zeros $f_1=f_2=0$ in $(\mathbb{C^*})^2$ is bounded
above by the mixed area of the two corresponding Newton polytopes
$\mathcal{M}(New(f_1),New(f_2))$}.
Here, $\mathbb{C^*} = \mathbb{C}\setminus\{0\}$.
To understand this theorem, one
should first associate a convex polytope to polynomial. A polytope
is a subset of $\mathbb{R}^n$ which is the convex hull of a finite
size of points. For example, in $\mathbb{R}^2$, the convex hull
$conv\{(0,0),(0,1),(1,0),(1,1)\}$ is a square. For a given
polynomial
\bea
f=c_1x^{\alpha_1}y^{\beta_1}+c_2x^{\alpha_2}y^{\beta_2}+\cdots+c_mx^{\alpha_m}y^{\beta_m}~,~~~\nonumber\eea
we can associate a Newton convex polytope
\bea New(f)=conv\{(\alpha_1,\beta_1),(\alpha_2,\beta_2),\ldots,
(\alpha_m,\beta_m)\}~.~~~\eea
Since $\alpha_i,\beta_i$'s are always non-negative integers, it is a
lattice convex polytope. Given two polytopes $P_1,P_2$, the
Minkowski sum is given by
\bea P_1+P_2=\{p_1+p_2: p_1\in P_1,p_2\in P_2\}~.~~~\eea
Then the mixed area $\mathcal{M}(P_1,P_2)$ is given by
\bea \mathcal{M}(P_1,P_2)=area(P_1+P_2)-area(P_1)-area(P_2)~.~~~\eea
We can apply the convex polytope method to the example polynomials
(\ref{PolyExample}), which is shown in Figure (\ref{polyesp}).
The mixed area is given by
\bea \mathcal{M}(New(f_1),New(f_2))={17\over 2}-{7\over
2}-1=4~.~~~\nonumber\eea
Following Bernstein's theorem, the number of zeros
for $f_1=f_2=0$ with general coefficients in $(\mathbb{C^*})^2$
is exactly $4$.  In case, the bound can be trivially
lifted from $(\mathbb{C^*})^2$ to $\mathbb{C}^2$. One remark is that, in
computing the mixed area, the two polynomials should be independent.
For example, two polynomials $f'_1=f_1$, $f'_2=f_2+y^3f_1$ have the
same zeros as $f_1=f_2=0$. However, if we include the vanishing term
$y^3f_1$ in $f'_2$ as vertices in polytope $New(f'_2)$, then we get
the wrong result. Before computing the area, we should remove the
redundant terms such as $y^3f_1$ in $f'_2$.

Bernstein's theorem can be generalized to higher dimension. {The
number of solutions in $(\mathbb{C^*})^n$ of $n$ polynomials in $n$
variables is bounded above by the mixed volume of $n$ Newton polytopes}. The
mixed volume of $P_1,P_2,\ldots, P_n$ in $\mathbb{R}^n$ is given by
formula
\bea \mathcal{M}(P_1,\ldots,P_n)=\sum_{J\subseteq
\{1,2,\ldots,n\}}(-1)^{n- n_J}\cdot volume(\sum_{j\in
J}P_j)~,~~~\eea
where $J$ is a non-empty subsets of $\{1,2,\ldots,n\}$ and
$n_J=1,2,\ldots, n$ is the length of $J$. The volume is Euclidean
volume in $\mathbb{R}^n$. For example, in $\mathbb{R}^4$, the mixed
volume is
\bea \mathcal{M}(P_1,P_2,P_3,P_4)&=&volume(P_1+P_2+P_3+P_4)-\sum_{1\leq i<j<k\leq 4}volume(P_i+P_j+P_k)\nonumber\\
&&+\sum_{1\leq i<j\leq 4}volume(P_i+P_j)-\sum_{1\leq i\leq 4}
volume(P_i)~.~~~\label{mixedvolume4d}\eea

\bibliographystyle{unsrt}
\bibliography{genus3Lv1}

\end{document}